\documentclass[nofootinbib, reprint, amsmath,amssymb, aps]{revtex4-1}
\usepackage{graphicx}
\usepackage{bm}
\graphicspath{ {../} }

\usepackage[dvipsnames]{xcolor}
\usepackage{footnote}
\usepackage{enumitem}
\usepackage{qcircuit}
\usepackage{float}
\usepackage{mathtools}
\usepackage{amsmath}
\usepackage{subcaption}
\usepackage[colorlinks]{hyperref}

\usepackage[bottom]{footmisc}

\usepackage{verbatim}

\usepackage{caption}

\begin{document}

\title{Molecular excited state calculations with the QEB-ADAPT-VQE}

\author{Yordan S. Yordanov$^{1,2}$}
\author{Crispin H. W. Barnes$^1$}
\author{David R. M. Arvidsson-Shukur$^{2, 1}$}

\affiliation{$\ ^1$ Cavendish Laboratory, Department of Physics, University of Cambridge, Cambridge CB3 0HE, United Kingdom}
\affiliation{$\ ^2$ Hitachi Cambridge Laboratory, J. J. Thomson Avenue, CB3 0HE,
Cambridge, United Kingdom}

\begin{abstract}

Calculations of molecular spectral properties, like photodissociation rates and absorption bands, rely on knowledge of the excited state energies of the molecule of interest.
Protocols based on the variational quantum eigensolver (VQE) are promising candidates to calculate such energies on emerging noisy intermediate scale quantum (NISQ) computers.
The successful implementation of these protocols on NISQ computers, relies on ans\"atze that can accurately approximate the molecular states and that can be implemented by shallow quantum circuits.
In this paper, we introduce the excited qubit-excitation-based adaptive (e-QEB-ADAPT)-VQE protocol to calculate molecular excited state energies. 
The e-QEB-ADAPT-VQE constructs efficient problem-tailored ans\"atze by iteratively appending evolutions of qubit excitation operators.
The e-QEB-ADAPT-VQE is an adaptation of the QEB-ADAPT-VQE protocol, which is designed to be independent on the choice of an initial reference state. 
We perform classical numerical simulations for LiH and BeH$_2$ to benchmark the performance of the e-QEB-ADAPT-VQE. 
We demonstrate that the e-QEB-ADAPT-VQE can construct highly accurate ans\"atze that require at least an order of magnitude fewer $CNOT$s than standard fixed UCC ans\"atze, such as the UCCSD and the GUCCSD.

\end{abstract}

\maketitle

\section{Introduction}

Quantum molecular simulations with the variational quantum eigensolver (VQE) \cite{vqe_general, vqe_general_2, vqe_2,VQE_accelerated, vqe_google_hf} are a promising application for emerging noisy intermediate-scale quantum (NISQ) \cite{NISQ,google_supreme, vqe_nisq} computers.
The VQE is a hybrid quantum-classical algorithm that utilizes the Rayleigh Ritz variational principle to determine the lowest eigenvalue of a Hamiltonian operator, by optimizing an ansatz. In particular the VQE can be used to solve the electronic structure problem \cite{CCSD}, and find the ground state configuration and energy of a molecule. By utilizing both a quantum and a classical computer, the VQE is less quantum hardware demanding
at the expense of requiring more quantum measurements and classical post-processing, as compared to purely quantum algorithms, like the Quantum
Phase Estimation (QPE) algorithm \cite{QPE}.
A major challenge for the practical realization of a molecular VQE simulation on NISQ computers is to construct a variationally flexible ansatz that: (1) accurately approximates the eigenstates of $H$; (2) is easy to optimize; and (3) can be implemented by a shallow circuit that uses few $2$-qubit entangling gates, e.g. $CNOT$s, which are the current bottleneck of NISQ computers.

The most widely used type of ans\"atze for molecular VQE simulations, are the unitary coupled cluster (UCC) ans\"atze \cite{UCCSD,UCCSD_2, k-upCCGSD, UCCSD_bogoliubov, optimized_uccsd, vqe_first_uccsd}. 
They were motivated by the classical coupled cluster theory \cite{CCSD}, and correspond to products of fermionic excitation evolutions.
Due to their fermionic structure, UCC ans\"{a}tze preserve many of the physical symmetries of electronic wavefunctions, which makes them accurate and easy to optimize at the same time. 
State of the art ADAPT-VQE protocols \cite{adapt_vqe, qubit_adapt_vqe, pruning_vqe, benchmark_adapt_vqe}, iteratively construct problem-tailored ans\"{a}tze, which 
consists of close to optimal number of fermionic excitation evolutions. These ans\"atze are implemented by shallow ansatz circuits, and have few variational parameters, while at the same time can be highly accurate.

In a previous work \cite{iqeb_vqe}, we introduced the qubit-excitation based (QEB)-ADAPT-VQE. 
Unlike the original fermionic-ADAPT-VQE \cite{adapt_vqe}, the QEB-ADAPT-VQE constructs a problem-tailored ansatz by appending ``qubit excitation evolutions'' (unitary evolutions of excitation operators that satisfy ``qubit commutation relations'' \cite{iqeb_vqe, parafermions, eff_circs, UCCSD_qubit}). Qubit excitation evolutions are implemented by simpler circuits than the standard fermionic excitation evolutions \cite{eff_circs}. Also, as demonstrated in Refs. \cite{qubit_adapt_vqe, iqeb_vqe, UCCSD_qubit}, both types of excitation evolutions approximate electronic wavefunctions comparably well. 
Thus, in Ref. \cite{iqeb_vqe} we demonstrated that the QEB-ADAPT-VQE can simulate molecular ground states, using shallower quantum circuits than the previous state of the art fermionic-ADAPT-VQE \cite{adapt_vqe} and qubit-ADAPT-VQE \cite{qubit_adapt_vqe} protocols.

In this work, we present a modified version of the QEB-ADAPT-VQE designed to simulate molecular excited states, which we call the excited(e)-QEB-ADAPT-VQE. The QEB-ADAPT-VQE, similarly to other VQE protocols, largely relies on an initial reference state, $|\psi_0\rangle$, that has a large overlap with the true ground state, $|E_0\rangle$. 
In particular, the QEB-ADAPT-VQE evaluates the energy gradients of individual qubit excitation evolutions in order to decide the best way to grow its ansatz at each iteration.
These energy gradients are evaluated for zero values of the variational parameters on the presumption that $|\psi_0\rangle$ is close to $|E_0\rangle$.
However, choosing an initial reference state that is guaranteed to be close an unknown excited state is challenging. 
We modify the ansatz growing strategy of the QEB-ADAPT-VQE, so that it does not require $|\psi_0\rangle$ to be close to the target (ground or excited) state. 
Additionally, we present a comparison of qubit and fermionic excitation evolutions in their ability to construct ans\"atze for excited states.

The paper is organized as follows:
In Sec. \ref{sec:theory} we include a theoretical introduction.
Section \ref{sec:algorithm}  describes the steps of the e-QEB-ADAPT-VQE protocol, and in Sec. \ref{sec:results} we benchmark its performance for the first excited states of LiH and BeH$_2$. 
Lastly, Sec. \ref{sec:qe_vs_fe_excited states} presents a comparison between qubit and fermionic excitation evolutions.

\section{Theory}\label{sec:theory}

\subsection{The electronic structure problem and the variational quantum eigensolver}\label{sec:vqe}

Finding the ground-state $|E_0\rangle$ and corresponding energy $E_0$ of a molecule is known as the ``electronic structure problem" \cite{CCSD}.
This problem can be solved by solving the time-independent Schr\"odinger equation $H |E_0\rangle = E_0 |E_0\rangle$, where $H$ is the electronic Hamiltonian of the molecule.
Within the Born-Oppenheimer approximation, where the nuclei of the molecule are assumed to be motionless, $H$ can be written in second quantized form as
\begin{equation}\label{eq:2nd_q_H}
{H} = \sum_{i,k}^{N_{_{MO}}} h_{i,k}^{\ } {a}_i^\dagger {a}_k^{\ } + \sum_{i,j,k,l}^{N_{_{MO}}} h_{i,j,k,l}^{\ } {a}_i^\dagger {a}_j^\dagger  {a}_k^{\ } {a}_l^{\ }.
\end{equation}
Here, $N_{_{MO}}$ is the number of considered spin-orbitals, ${a}^\dagger_i$ and ${a}_i$ are the fermionic ladder operators, corresponding to the $i^{th}$ molecular spin-orbital, and the factors $h_{ij}$ and $h_{ijkl}$ are one- and two-electron integrals, written in a spin-orbital basis \cite{CCSD}.
The Hamiltonian expression in Eq. \ref{eq:2nd_q_H} can be mapped to quantum-gate operators using a qubit encoding method, e.g. the Jordan-Wigner (JW) \cite{JW_encode} or the Bravyi-Kitaev \cite{BK_encode} methods. 
Throughout this work, we assume the more straightforward JW encoding, where the occupancy of the $i^{th}$ molecular spin-orbital is represented by the state of the $i^{th}$ qubit.

The fermionic ladder operators ${a}^\dagger_i$ and ${a}_i$ satisfy anti-commutation relations
\begin{equation}\label{eq:anti_comm}
\{a_i,a^\dagger_j\} = \delta_{i,j}, \ \ \{a_i, a_j\} = \{a_i^\dagger, a^\dagger_j\} = 0.
\end{equation}
Within the JW encoding, ${a}^\dagger_i$ and ${a}_i$ can be written in terms of quantum gate operators as
\begin{equation}\label{eq:a*}
a_i^\dagger = Q_i^\dagger \prod_{r=0}^{i-1} Z_r = \frac{1}{2}(X_i-iY_i)\prod_{r=0}^{i-1} Z_r \; \; \mathrm{and} 
\end{equation}
\begin{equation}\label{eq:a}
a_i =  Q_i \prod_{r=0}^{i-1} Z_r=\frac{1}{2}(X_i+iY_i)\prod_{r=0}^{i-1} Z_r,
\end{equation}
where 
\begin{equation}\label{eq:Q*}
Q_i^\dagger \equiv \frac{1}{2}(X_i - iY_i)  \; \; \mathrm{and} \ Q_i \equiv \frac{1}{2}(X_i + iY_i).
\end{equation}
We refer to ${Q_i^\dagger}$ and ${Q_i}$ as qubit creation and annihilation operators, respectively.
The operators $Q^\dagger_i$ and $Q_i$ satisfy the ``qubit commutation relations'' \cite{parafermions, eff_circs, iqeb_vqe}
\begin{align}\label{eq:Q_Q}
 \{Q_i, Q_i^\dagger\} = I,  [Q_i, Q^\dagger_j] = 0  \text{ if } \ i \neq j , \; \;  \mathrm{ and } \nonumber \\ \; \; [Q_i, Q_j] = [Q_i^\dagger, Q^\dagger_j] = 0 \text{ for all }i,j.  
\end{align}
 We will use ${Q_i^\dagger}$ and ${Q_i}$ in Sec. \ref{sec:q_exc} to define qubit excitation evolutions.

Substituting Eqs. \eqref{eq:a*} and \eqref{eq:a} into Eq. \eqref{eq:2nd_q_H}, $H$ can be written as 
\begin{equation}\label{eq:p_H}
{H} =  \sum_r h_r \prod_{s=0}^{N_{MO}-1} \sigma_s^r,
\end{equation}
where $\sigma_s$ is a Pauli operator ($X_s$, $Y_s$, $Z_s$ or $I_s$) acting on qubit $s$, and $h_r$ (not to be confused with $h_{ik}$ and $h_{ijkl}$) is a real scalar coefficient{\color{blue}}. 
The number of terms in Eq. \eqref{eq:p_H} scales as $O(N_{_{MO}}^4)$. 
The expectation value of $H$ can be evaluated by individually measuring on a quantum computer the expectation values of all Pauli strings in Eq. \eqref{eq:p_H}.

Once $H$ is mapped to a Pauli string representation, the VQE can be used to minimize the expectation value $E(\pmb{\theta})=\langle \psi(\pmb{\theta})|H|\psi(\pmb{\theta})\rangle$, where $|\psi(\pmb{\theta})\rangle$ is a trial state.
The VQE relies upon the Rayleigh-Ritz variational principle
\begin{equation}\label{eq:R_R}
\langle \psi(\pmb{\theta})|H|\psi(\pmb{\theta})\rangle \geq E_0 ,
\end{equation}
to find an lower-bounded estimate for $E_0$. 
The VQE is a hybrid-quantum-classical algorithm that uses a quantum computer to prepare the trial state $|\psi (\pmb{\theta})\rangle $ and evaluate $E(\pmb{\theta})$, and a classical computer to process the measurement data and update  $\pmb{\theta}$ at each iteration.
The trial state $|\psi (\pmb{\theta})\rangle = U(\pmb{\theta}) |\psi_0\rangle$ is generated by the ansatz $U(\pmb{\theta})$ applied to an initial reference state $|\psi_0\rangle$.

\subsection{Qubit and fermionic excitation evolutions}\label{sec:q_exc}

The most widely used type of ans\"atze for molecular VQE simulations, are the unitary coupled cluster (UCC) ans\"atze, motivated by the classical coupled cluster method \cite{CCSD}. UCC ans\"atze are constructed as products of fermionic excitation evolutions.
The QEB-ADAPT-VQE and the e-QEB-ADAPT-VQE instead construct ans\"atze that consist of qubit excitation evolutions. 
This subsection provides definitions of the two types of unitary operations. 

\subsubsection{Fermionic excitation evolutions}

Single and double fermionic excitation operators, are defined, respectively, by the skew-Hermitian operators
\begin{equation}\label{eq:s_f_exc_op_aa}
T_{ik} \equiv  a^\dagger_i a_k - a^\dagger_k a_i \; \; \mathrm{and}
\end{equation}
\begin{equation}\label{eq:d_f_exc_op_aa}
T_{ijkl} \equiv  a^\dagger_i a^\dagger_j a_k a_l - a^\dagger_k a^\dagger_l a_i a_j.
\end{equation}
Single and double fermionic excitation evolutions are thus given, respectively, by the unitary operators
\begin{equation}\label{eq:s_f_exc}
A_{ik}(\theta) = e^{\theta  T_{ik} } = \exp \left[ \theta(a^\dagger_i a_k - a^\dagger_k a_i)\right] \; \; \mathrm{and}
\end{equation}
\begin{equation}\label{eq:d_f_exc}
A_{ijkl}(\theta) = e^{ \theta T_{ijkl} } = \exp \left[ \theta( a^\dagger_i a^\dagger_j a_k a_l - a^\dagger_k a^\dagger_l a_i a_j)  \right].
\end{equation}
Using equations \eqref{eq:a*} and \eqref{eq:a}, for $i<j<k<l$, $A_{ik}$ and $A_{ijkl}$ can be expressed in terms of quantum gate operators as
\begin{equation}\label{eq:s_f_exc_qo}
A_{ij}(\theta)=\exp\left[i\frac{\theta}{2}(X_i Y_k - Y_i X_k)\prod_{r=i+1}^{k-1}Z_r \right] \; \; \mathrm{and}
\end{equation}
\begin{align}
\label{eq:d_f_exc_qo}
A_{ijkl}(\theta) =\exp \Bigg[ i\frac{\theta}{8} (X_i Y_j X_k X_l + Y_i X_j X_k X_l + Y_i Y_j Y_k X_l     \nonumber  \\ + Y_i Y_j X_k Y_l  - X_i X_j Y_k X_l  - X_i X_j X_k Y_l  \nonumber  \\ - Y_i X_j Y_k Y_l  - X_i Y_j Y_k Y_l ) \prod_{r=i+1}^{j-1} Z_r\prod_{r'=k+1}^{l-1} Z_{r'} \Bigg].
\end{align}
$CNOT$-efficient circuits to implemented single and double fermionic excitation evolutions were derived in Ref. \cite{eff_circs}.

\subsubsection{Qubit excitation evolutions}

Qubit excitation evolutions are generated by ``qubit excitation operators''.
Single and double qubit excitation operators are defined, respectively, by the skew-Hermitian operators
\begin{equation}\label{eq:s_q_exc}
\tilde{T}_{ik} =  Q^\dagger_i Q_k - Q^\dagger_k Q_i \; \; \mathrm{and}
\end{equation}
\begin{equation}\label{eq:d_q_exc}
\tilde{T}_{ijkl} = Q^\dagger_i Q^\dagger_j Q_k Q_l - Q^\dagger_k Q^\dagger_l Q_i Q_j.
\end{equation}
Hence, single and double qubit excitation evolutions are defined, respectively, by the unitary operators
\begin{equation}
\tilde{A}_{ik}(\theta) = e^{\theta \tilde{T}_{ik}} =  \exp \big[ \theta(Q^\dagger_i Q_k - Q^\dagger_k Q_i) \big] \; \; \mathrm{and}
\end{equation}
\begin{equation}
\tilde{A}_{ijkl}(\theta) = e^{\theta \tilde{T}_{ijkl}} = \exp \big[ \theta(Q^\dagger_i Q^\dagger_j Q_k Q_l - Q^\dagger_k Q^\dagger_l Q_i Q_j) \big].
\end{equation}
Using Eq. \eqref{eq:Q*}, $\tilde{A}_{ik}$ and $\tilde{A}_{ijkl}$ can be re-expressed in terms of quantum gate operators:
\begin{equation}\label{eq:s_q_exc_qo}
\tilde{A}_{ik}(\theta)= \exp \Big[ i\frac{\theta}{2}(X_i Y_k - Y_i X_k) \Big] \; \; \mathrm{and}
\end{equation}
\begin{align}
\label{eq:d_q_exc_qo}
\tilde{A}_{ijkl}(\theta) = \exp \Big[ i\frac{\theta}{8} (X_i Y_j X_k X_l + Y_i X_j X_k X_l \nonumber \\ + Y_i Y_j Y_k X_l    +  Y_i Y_j X_k Y_l  - X_i X_j Y_k X_l \nonumber \\ - X_i X_j X_k Y_l  - Y_i X_j Y_k Y_l - X_i Y_j Y_k Y_l ) \Big] .
\end{align}
$CNOT$-efficient circuits to implement single and double qubit excitation evolutions were derived in Ref. \cite{eff_circs}.

\subsection{Finding excited state energies with the VQE}\label{sec:overlap}

The two most common methods to calculate excited state energies with the VQE are the quantum subspace expansion \cite{subspace_expansion_2, subspace_expansion_3, subspace_expansion_1}, and the overlap-based method \cite{overlap_method_1, overlap_method_2}.
In this work we use the latter overlap-based method.

As the name suggest the overlap-based method works by including the overlap between previously found, eigenstates of $H$ and the state we are currently searching for in the cost function of the VQE.
For example, after we find the ground state $|E_0\rangle$ we modify H as
\begin{equation}\label{eq:h_1}
H \rightarrow H_1 = H + \alpha_0 |E_0\rangle \langle E_0|,
\end{equation}
where $\alpha_0$ is a real positive scalar coefficient. If $\alpha_0 + E_0 > E_1$ \footnote{To make sure that $\alpha_0 + E_0 > E_1$, prior to knowing $E_1$, we can choose an arbitrary value for $\alpha_0$, which is large compared to the energy scale of the problem.}, where $E_1$ is the energy of the first excited state of $H$, the lowest energy eigenvalue of the modified Hamiltonian $H_1$ will be shifted to $E_1$. Hence, running the VQE for $H_1$ should output an estimate for $E_1$:
\begin{multline}
E_1 = \min_{\pmb{\theta_1}} \langle \psi_0 | U_1^\dagger(\pmb{\theta_1}) H_1 U_1(\pmb{\theta_1}) | \psi_0 \rangle = \\
\min_{\pmb{\theta_1}} \langle \psi_0 | U_1^\dagger(\pmb{\theta_1}) \big(H + \alpha_0 |E_0\rangle \langle E_0|\big) U_1(\pmb{\theta_1}) | \psi_0 \rangle
\end{multline}
After $E_1$ is found the same procedure can be repeated recursively multiple times to find the next energy eigenvalues of $H$. The $k^\text{th}$ energy eigenvalue of $H$, $E_k$, can be estimated as
\begin{multline}\label{eq:E_k}
E_1 = \min_{\pmb{\theta_k}} \langle \psi_0 | U_k^\dagger(\pmb{\theta_k}) H_k U_k(\pmb{\theta_k}) | \psi_0 \rangle = \\
\min_{\pmb{\theta_k}} \langle \psi_0 | U_k^\dagger(\pmb{\theta_k}) \big(H + \sum_{r=0}^{k-1} \alpha_r |E_r\rangle \langle E_r |\big) U_k(\pmb{\theta_k}) | \psi_0 \rangle
\end{multline}
The first term in Eq. \eqref{eq:E_k}, is calculated as described above, by measuring the expectation values of the Pauli string terms in the expression for $H$ [Eq. \eqref{eq:p_H}].
The overlap terms in Eq. \eqref{eq:E_k} can be calculated with the SWAP test \cite{swap_test}.

\section{The excited-QEB-ADAPT-VQE}\label{sec:algorithm}

The e-QEB-ADAPT-VQE algorithm finds an estimate for the $k^{\text{th}}$ excited state of an electronic Hamiltonian, $H$, by constructing a problem-tailored ansatz.
The ansatz is constructed by iteratively appending qubit excitation evolutions, based on the greedy strategy to obtain the lowest estimate for the energy, $E(\pmb{\theta})$, at each iteration.
In this subsection, we describe the preparation components, and the iterative ansatz-constructing loop, of the e-QEB-ADAPT-VQE algorithm. For a description of the original QEB-ADAPT-VQE, we refer the reader to Ref. \cite{iqeb_vqe}.

 First, we transform the $H$ to a quantum-gate-operator representation as described in  Sec. \ref{sec:vqe}.
This involves the calculation of the one- and two-electron integrals $h_{ik}$ and $h_{ijkl}$ [Eq. \eqref{eq:2nd_q_H}], which can be done efficiently (in time polynomial in $N_{MO}$) on a classical computer.
If we want to find the energy of the excited state $|E_k\rangle$, $E_k$, we add the additional overlap terms to $H$, as described in Sec. \ref{sec:overlap}, to get $H_k$.

Second, we define an ansatz element pool $\mathbb{P}(\tilde{A}, N_{MO})$ of all unique single and double qubit excitations, $\tilde{A}_{ik}(\theta)$ and $\tilde{A}_{ijkl}(\theta)$, respectively, for $i,j,k,l \in \{0,N_{MO}-1\}$.
The size of this pool is $| \mathbb{P}(\tilde{A}, N_{MO}) |=\binom{N_{MO}}{2} + 3 \binom{N_{MO}}{4}$. 

Third, we choose an initial reference state $|\psi_0\rangle$. 
As we mentioned in the introduction, the performance of the e-QEB-ADAPT-VQE is intended to be independent of the choice of $|\psi_0\rangle$. Hence, we arbitrarily choose the initial reference state to be the Hartree-Fock state, but in principle any computational basis state with a Hamming weight equal to the number of electrons, $N_e$, will do.

Now we are ready to begin constructing the ansatz. We set the iteration number to $m=1$, the ansatz to the identity $U \rightarrow U^{(0)} = I$, and initiate the iterative ansatz-constructing loop. We describe the five steps of the $m^{th}$ iteration below. Afterwards we comment on these steps.

\begin{enumerate}

\item Prepare trial state $|\psi^{[m-1]}\rangle = U\left(\pmb{\theta}^{[m-1]}\right) |\psi_0\rangle$, with  values for $\pmb{\theta}^{[m-1]}$ as determined in the previous iteration.

\item For each (single or double) qubit excitation evolution $\tilde{A_p}(\theta_p) \in \mathbb{P}(\tilde{A}, N_{MO})$: 
\begin{enumerate}
\item Run the VQE to find 

$$\min\limits_{\theta_p} E(\theta_p)=  \min\limits_{\theta_p} \langle \psi^{[m-1]}|  \tilde{A_p}^\dagger(\theta_p) H_k \tilde{A_p}(\theta_p)| \psi^{[m-1]} \rangle. \nonumber$$

\item Calculate the energy reduction magnitude $ E^{[m-1]} - \min\limits_{\theta_p} E(\theta_p) $

\end{enumerate}

\item Identify the set of $n$ qubit excitation evolutions, $\mathbb{\tilde{A}}^{[m]}(n)$, corresponding to the $n$ largest energy reductions in the previous step.
For $\tilde{A_p}(\theta_p) \in  \mathbb{\tilde{A}}^{[m]}(n)$:
\begin{enumerate}
\item Run the VQE to find $\min\limits_{\pmb{\theta}^{[m-1]},\theta_p} E\left(\pmb{\theta}^{[m-1]},\theta_p\right)=$
\begin{equation}
 \min\limits_{\pmb{\theta}^{[m-1]},\theta_p} \langle \psi_0| U^\dagger\big(\pmb{\theta}^{[m-1]}\big) \tilde{A_p}^\dagger(\theta_p) H_k \tilde{A_p}(\theta_p)
U\big(\pmb{\theta}^{[m-1]}\big) | \psi_0 \rangle. \nonumber
\end{equation}

\item Calculate the energy reduction $\Delta {E}^{[m]}_p = E^{[m-1]} - \min\limits_{\pmb{\theta}^{[m-1]},\theta_p} E\left(\pmb{\theta}^{[m-1]},\theta_p\right) $.

\item Save the (re)optimized values of $\pmb{\theta}^{[m-1]} \cup  \{ \theta_p \} $ as $\pmb{\theta}^{[m]}_p$.

\end{enumerate}

\item Identify the qubit excitation evolution $\tilde{A}^{[m]}\left(\theta^{[m]}\right) \equiv \tilde{A}_{p'}(\theta_{p'})$ corresponding to the largest energy reduction $\Delta {E}^{[m]} \equiv \Delta {E}_{p'}^{[m]}$ in the previous step.

If $\Delta E^{[m]} < \epsilon $, where $\epsilon>0$ is an energy threshold:
\begin{enumerate}
\item Exit
\end{enumerate}
Else:
\begin{enumerate}
\item  Append $\tilde{A}^{[m]}\left(\theta^{[m]}\right)$ to the ansatz:
\begin{equation}
U\big(\pmb{\theta}^{[m-1]}\big) \rightarrow U\big(\pmb{\theta}^{[m]}\big) = \tilde{A}^{[m]}\big(\theta^{[m]}\big) U\big(\pmb{\theta}^{[m-1]}\big)
\end{equation}
\item Set $E^{[m]} = E^{[m-1]} - \Delta {E}^{[m]}_p$

\item Set the values of the new set of variational parameters, $\pmb{\theta}^{[m]} = \pmb{\theta}^{[m-1]} \cup  \{ \theta_{p'} \}$, to $\pmb{\theta}^{[m]}_{p'}$

\end{enumerate}

\item Enter the $m+1$ iteration by returning to step $1$

\end{enumerate}

We now elaborate on the steps outlined above.

The $m^{\text{th}}$ iteration of  the e-QEB-ADAPT-VQE starts by preparing the trial state $|\psi^{[m-1]}\rangle$ obtained in the $(m-1)^{\text{th}}$ iteration.

At each iteration we aim to identify the qubit excitation evolution that, when appended to the ansatz, would maximise the energy reduction $E(\pmb{\theta}^{[m-1]}) - E(\pmb{\theta}^{[m]})$.
To identify such a qubit excitation evolution, first we calculate (step 2) the individual energy reduction contributions of each qubit excitation evolution $\tilde{A}_p(\theta_p) \in \mathbb{P}(\tilde{A}, N_{MO})$. 
Each of these energy reductions is calculated by a single-parameter VQE optimization performed to minimize the energy expectation value $ \langle \psi^{[m-1]} |\tilde{A}_p^\dagger (\theta_p)H  \tilde{A}_p(\theta_p) |\psi^{[m-1]}\rangle$. 
In this way, we get an indication by how much the energy expectation value will be reduced when each $\tilde{A}_p(\theta_p)$ is appended to the ansatz $U(\pmb{\theta}^{[m-1]})$ and the energy is minimized along the full set of parameters $\pmb{\theta}^{[m-1]} \cup \{ \theta_p\}$, at a reduced cost of performing $O(N_{MO}^4)$ single-parameter VQE optimizations (instead of $O(N_{MO}^4)$ $m$-parameter VQE optimizations).

In the original QEB-ADAPT-VQE such an indication is obtained by measuring the energy gradients $\left\{ \frac{\partial}{\partial \theta_p} \langle \psi^{[m-1]} |\tilde{A}_p^\dagger(\theta_p) H  \tilde{A}_p(\theta_p) |\psi^{[m-1]}\rangle \right\}$. 
However, these energy gradients must be evaluated for some values of the respective parameters $\left\{ \theta_p\right\}$. If we are approximating the ground state $|E_0\rangle$ and we have a high level of confidence that $|\psi_0\rangle$ has a large overlap with $|E_0\rangle$, then the gradients can be evaluated conveniently for $\left\{ \theta_p=0\right\}$. But if we are approximating the excited state $|E_k\rangle$, where we do not know how close $|\psi_0\rangle$ and $|E_k\rangle$ are, measuring the energy gradients for $\left\{ \theta_p=0\right\}$, or any other arbitrary values of $\{\theta_p\}$, can be a poor indicator, leading to an inefficient and slow construction of the ansatz, and a possibility of getting stuck in local energy minimum (see  Sec. \ref{sec:exc_iqeb_vs_iqeb}).
Instead the e-QEB-ADAPT-VQE measures the individual energy reduction contribution for each qubit excitation evolution, which can be obtained by a single-parameter VQE optimization, using a direct search minimizer \cite{direct_search}, e.g. the Nelder Mead \cite{nelder_mead}, irrespectively of the overlap of $|\psi_0\rangle$ and $|E_k\rangle$.

If we denote by $N_H$ the number of terms in the Pauli string representation of $H$ [Eq. \eqref{eq:p_H}], calculating a single energy-gradient requires measuring $2 N_H$ expectation values. On the other hand, running a single-parameter VQE requires measuring $\gamma N_H$ expectation values, where $\gamma$ is the number of function evaluations required for a single parameter minimization.
Hence, the additional cost of the technique pursued here in comparison to using energy gradients as in the original QEB-ADAPT-VQE, is a factor of $\gamma/2$ more quantum computer measurements in step 2\footnote{For the Nelder Mead $\gamma$ is on the order of $10$.}.
In addition, once the ansatz reaches some critical size, so that $|\psi^{[m-1]}\rangle$ can be assumed to have a significant overlap with $|E_k\rangle$, we can switch to the cheaper QEB-ADAPT-VQE.

The individual energy reductions calculated in step 2, indicate how much each qubit excitation evolution can decrease $E^{(m-1)}$ when appended to $U(\pmb{\theta}^{(m-1)})$. 
However, the largest individual energy reduction does not necessarily correspond to the largest energy reduction when the ansatz is optimized over all variational parameters. 
In step 3, we identify the set of $n$ qubit excitation evolutions with the individual energy reductions: $\mathbb{\tilde{A}}^{[m]}(n)$. We assume that $\mathbb{\tilde{A}}^{[m]}(n)$ likely contains the qubit excitation evolution that reduces $E^{[m-1]}$ the most. For each of the $n$ qubit excitation evolutions in $\mathbb{\tilde{A}}^{[m]}(n)$, we run the VQE, for all variational parameters, with the ansatz from the previous iteration to calculate how much it  contributes to the energy reduction. 
If multiple quantum devices are available, step 3 can be parallelized.

In step 4, we pick the qubit excitation evolution, $\tilde{A}^{[m]}\left(\theta^{[m]}\right)$, corresponding to the largest energy reduction, $\Delta {E}^{[m]}$. If $\Delta {E}^{[m]}$ is below some threshold $\epsilon > 0$, we exit the iterative loop.
If instead the $|\Delta {E}^{[m]}|> \epsilon$, we add $\tilde{A}^{[m]}\left(\theta^{[m]}\right)$ to the ansatz and begin the next iteration.

\onecolumngrid

\begin{figure}

\includegraphics[width=16cm]{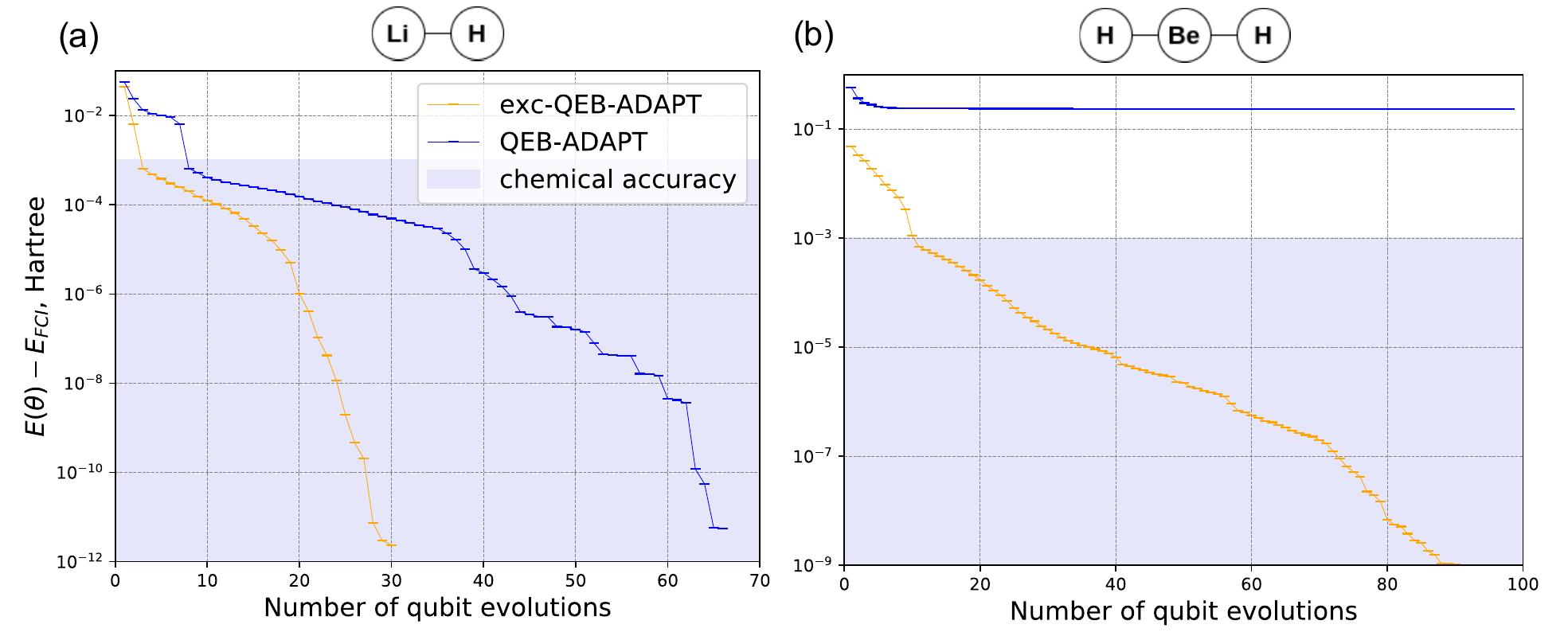}

\caption{Energy convergence plots for the first excited states of LiH and BeH$_2$ in the STO-3G basis at equilibrium bond distances of $r_{\text{Li-H}}=1.546 {\textup{\AA}}$ and $r_{\text{Be-H}}=1.316 {\textup{\AA}}$, respectively. The plots are obtained with the QEB-ADAPT-VQE and the e-QEB-ADAPT-VQE for $n=10$.}
\label{fig:iqeb_vs_exc_iqeb}
\end{figure}

\twocolumngrid

\section{Benchmarking the e-QEB-ADAPT-VQE}\label{sec:results}

In this section, we benchmark the performance of the e-QEB-ADAPT-VQE by finding the excited state energies of LiH and BeH$_2$, which have been simulated on real quantum computers \cite{vqe_first_uccsd, UCCSD}. 
The results presented here are based on classical numerical VQE simulations, performed with a custom in-house code\footnote{The code and the numerical data are available upon request from the authors. }. 
The Hamiltonians for LiH and BeH$_2$ are represented in the STO-3G orbital basis set \cite{sto_3g}, with no frozen orbitals assumed. Hence, the wavefunction of LiH is represented by a $12$ qubit state, and that of BeH$_2$ by a $14$ qubit state.
In step 2 of the e-QEB-ADAPT-VQE, the VQE optimization is performed with the direct-search Nelder Mead method \cite{nelder_mead}, and in step 3 with the gradient descent BFGS method \cite{bfgs} .

\subsection{The QEB-ADAPT-VQE vs the e-QEB-ADAPT-VQE}\label{sec:exc_iqeb_vs_iqeb}

In this subsection we compare the convergence rates, to the first excited state energies of LiH and BeH$_2$, of the QEB-ADAPT-VQE and the e-QEB-ADAPT-VQE.
With this comparison we demonstrate that the energy-gradient-based ansatz-growing strategy of the QEB-ADAPT-VQE is unsuitable for excited states.

Figure \ref{fig:iqeb_vs_exc_iqeb} presents energy convergence plots for the first excited states of LiH and BeH$_2$ in the STO-3G basis at equilibrium bond distances of $r_{\text{Li-H}}=1.546 {\textup{\AA}}$ and $r_{\text{Be-H}}=1.316 {\textup{\AA}}$, respectively. The plots are obtained with each of the two protocols.
In the case of LiH the QEB-ADAPT-VQE converges slower, requiring more than twice as many ansatz-constructing iterations, than the e-QEB-ADAPT-VQE. Since each ansatz-constructing iteration corresponds to a qubit excitation evolution in the ansatz, the ansatz constructed by the QEB-ADAPT-VQE is roughly twice as large as the one constructed by the e-QEB-ADAPT-VQE.
As suggested in Sec. \ref{sec:algorithm} above, the underlying reason for this is the inability of the energy-gradient-based ansatz-growing strategy of the QEB-ADAPT-VQE to grow an efficient ansatz when the initial reference state, $|\psi_0\rangle$, does not have a significant overlap with the target excited state, $|E_k\rangle$. 
In the case of BeH$_2$, the QEB-ADAPT-VQE completely fails to converge to chemical accuracy, likely getting stuck in a local energy minimum.
Nevertheless, the e-QEB-ADAPT-VQE, being independent on the overlap of $|\psi_0\rangle$ and $|E_k\rangle$, converges successfully for both molecules.

\subsection{Energy dissociation curves}\label{sec:benchmark_excited_iqeb}

In this section, we benchmark the performance of the e-QEB-ADAPT-VQE, for $\epsilon=10^{-6}$ Hartree and $\epsilon=10^{-8}$ Hartree, by obtaining energy dissociation plots for the first excited states of LiH and BeH$_2$.
For a comparison, we also include energy dissociation plots obtained with the VQE using the standard UCCSD \cite{UCCSD, UCCSD_2} ansatz, which consists of single and double fermionic excitation evolutions above the Hartree-Fock state, and the larger generalized UCCSD (GUCCSD) ansatz \cite{k-upCCGSD}, which consists of all unique single and double fermionic excitation evolutions.

\begin{figure}[H]
  \centering
   \includegraphics[width=13cm]{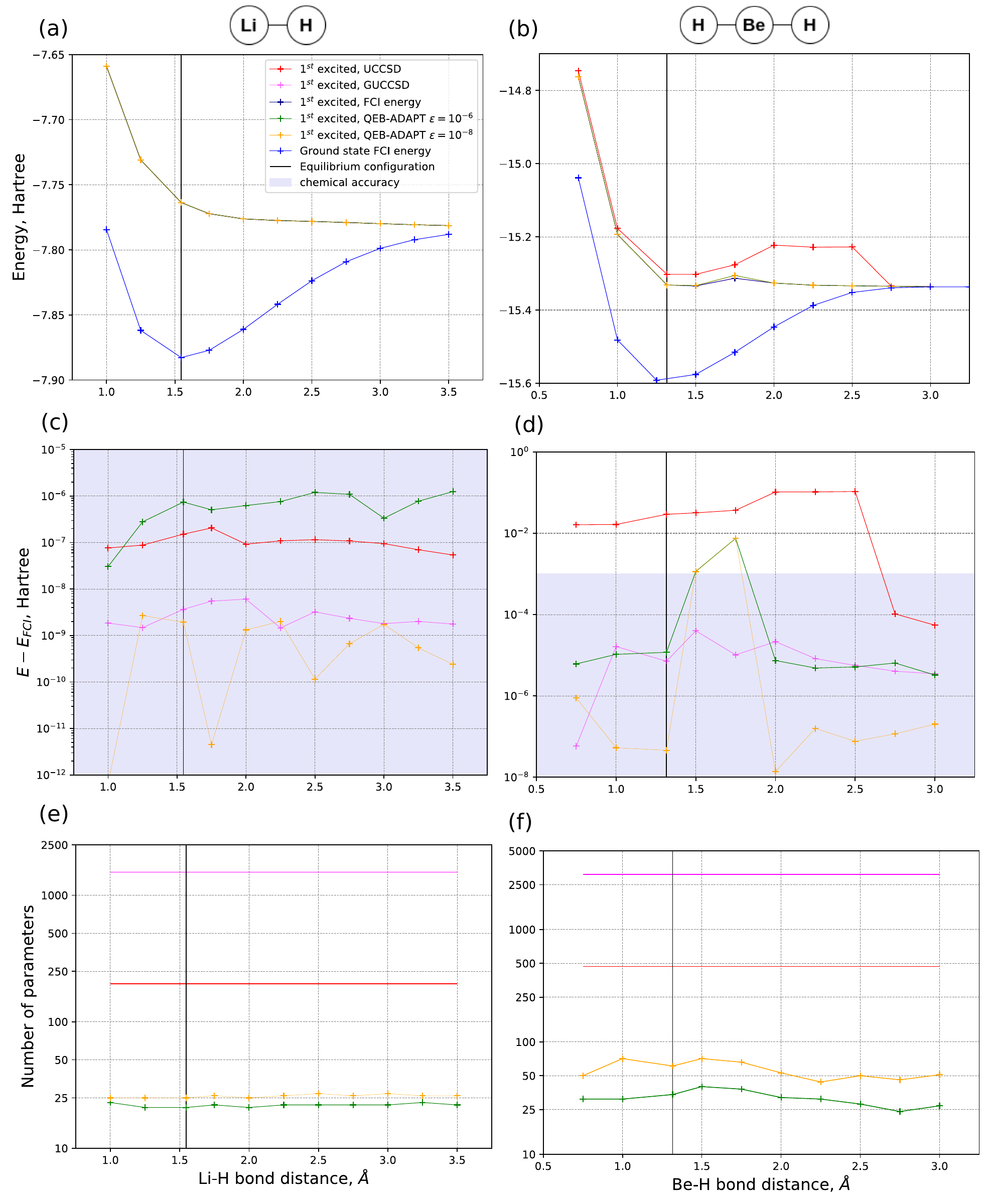}

    \caption[Energy dissociation plots for the first excited states of LiH and BeH$_2$]{ Energy dissociation curves for $\text{LiH}$ and $\text{BeH}_2$ in the STO-3G orbital basis. \textbf{a,b} Absolute values for the estimated first excited state energies. \textbf{c,d} Error in the estimated energy values with respect to the exact FCI energy. \textbf{e,f} Number of variational parameters, also equal to the number of qubit/fermionic evolutions, of the ansatz used by each method.
}
\label{fig:exc_dissociation}
\end{figure}

The energy dissociation curves for the two molecules are given in Figs. \ref{fig:exc_dissociation}a and \ref{fig:exc_dissociation}b.
Figures \ref{fig:exc_dissociation}c and \ref{fig:exc_dissociation}d show the errors of each method with respect to the full configuration interaction (FCI) energy \cite{CCSD, classical_methods}.
In the case of LiH, all methods achieve chemical accuracy.
Figures \ref{fig:exc_dissociation}e and \ref{fig:exc_dissociation}f show the numbers of variational parameters, which are also equivalent to the number of qubit/fermionic excitation evolutions, used in the ansatz of each method.
The ans\"atze constructed by the e-QEB-ADAPT-VQE, for LiH, are extremely compact, consisting of at most $27$ qubit evolutions, whereas the UCCSD and the GUCCSD consist of $200$ and $1521$ fermionic evolutions, respectively.

In the case of BeH$_2$ the results are more complicated. First, the UCCSD-VQE fails to achieve chemical accuracy for the majority of bond distances (Fig. \ref{fig:exc_dissociation}d).
This result is not surprising as the simple UCCSD ansatz is not suitable to approximate strongly correlated states such as excited states.
Second, in Fig. \ref{fig:exc_dissociation}d we see that the e-QEB-ADAPT-VQE fails to achieve chemical accuracy for bond distances of $r_{Be-H}=1.5 \textup{\AA}$ and $r_{Be-H}=1.75 \textup{\AA}$.
We comment on the reason for this in Sec. \ref{sec:fail}.
The only method that achieves chemical accuracy at all bond distances is the GUCCSD-VQE, because of the highly variationally flexible GUCCSD ansatz.
Nevertheless, for the majority of bond distances the e-QEB-ADAPT-VQE constructs ans\"atze that are more accurate and consist of nearly $50$ times fewer ansatz elements than the GUCCSD (Figs. \ref{fig:exc_dissociation}d and \ref{fig:exc_dissociation}f).

These results together with the fact that qubit evolutions are implemented by simpler circuits than fermionic evolutions, implies that the ans\"atze constructed by the e-QEB-ADAPT-VQE are implemented by much shallower circuits that have much fewer $CNOTs$ than the UCCSD and the GUCCSD ans\"atze.
Table \ref{table:cnots} summarizes the $CNOT$ counts for the UCCSD and the GUCCSD ans\"atze, and the ans\"atze constructed by the e-QEB-ADAPT-VQE($\epsilon=10^{-8}$) for each molecule. For the e-QEB-ADAPT-VQE($\epsilon=10^{-8}$), the maximum $CNOT$ counts over all bond distances for each molecule are shown.
As we can see in the table, the ans\"atze constructed by the e-QEB-ADAPT-VQE use at least $10$ times fewer $CNOT$s than the UCCSD, and more than $60$ times fewer $CNOT$s than the GUCCSD.

\begin{table}[H]
\centering
\begin{tabular}{c | c | c | c}
 & UCCSD & GUCCSD & e-QEB-ADAPT-VQE \\
 \hline
LiH & 3496 & 29447 & 311 \\
BeH$_2$ & 8980 & 64064 & 896
\end{tabular}
\caption{$CNOT$ counts for the UCCSD, the GUCCSD, and the ans\"atze constructed by the e-QEB-ADAPT-VQE($\epsilon=10^{-8}$) for the first excited states of LiH and BeH$_2$ in the STO-3G basis. }
\label{table:cnots}
\end{table}

\onecolumngrid

\twocolumngrid

\subsection{Finding the wrong excited state energy}\label{sec:fail}

\begin{figure}[H]
\centering
\includegraphics[width=14cm]{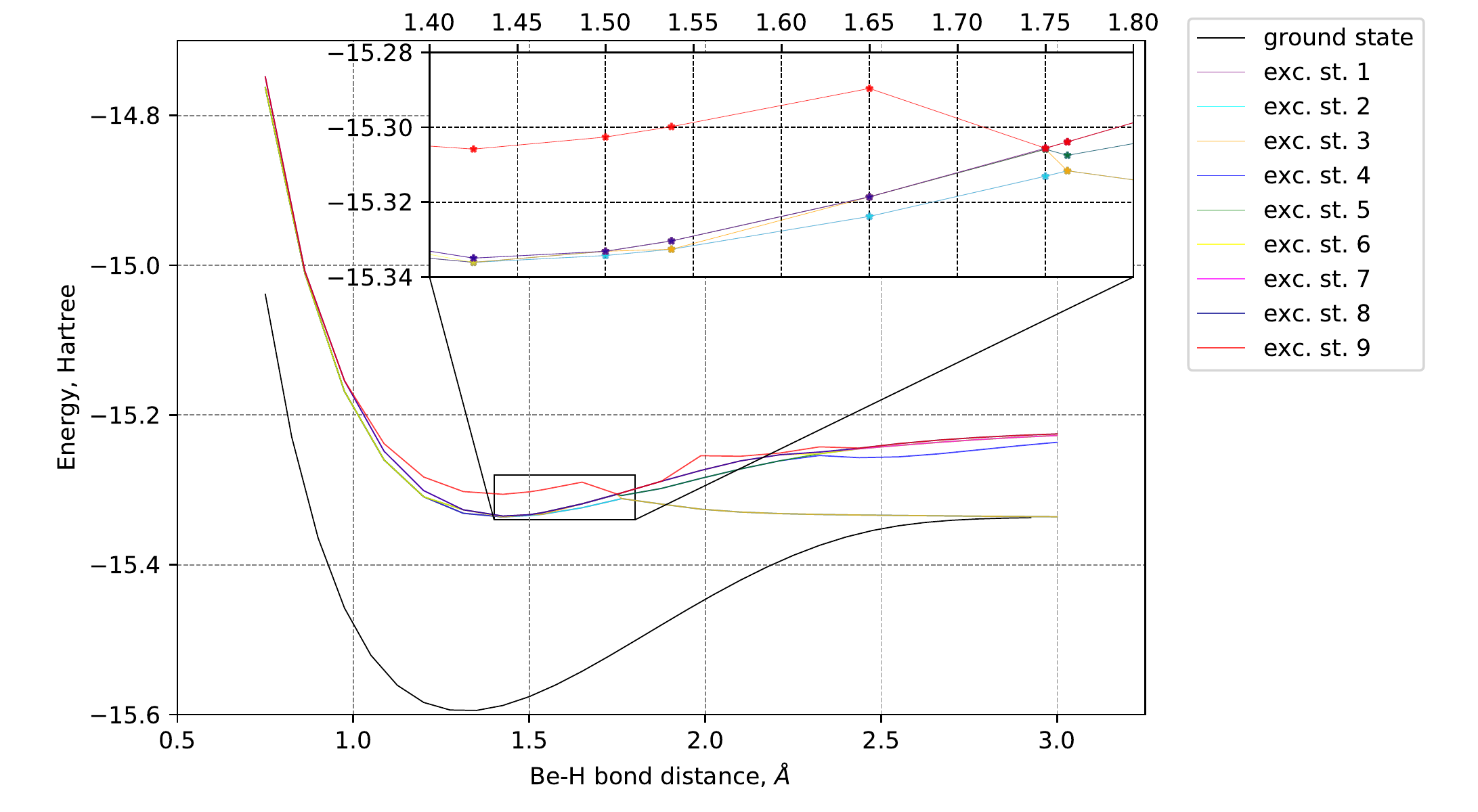}
\caption[Excited state energies for BeH$_2$]{FCI energies for the $10$ lowest energy states of BeH$_2$ in the STO-3G basis. The numbers of the excited states correspond to their order at the ground state equilibrium bond distance of $r_{Be-H}=1.316 \AA$.}
\label{fig:beh2_exc_fci_energies}
\end{figure}

As we saw in Fig. \ref{fig:exc_dissociation}d, the e-QEB-ADAPT-VQE fails to achieve chemical accuracy for BeH$_2$ at bond distances $r_{Be-H}=1.5 \textup{\AA}$ and $r_{Be-H}=1.75 \textup{\AA}$.
In order to investigate the reason for this see Fig. \ref{fig:beh2_exc_fci_energies}, which depicts the FCI energies for the $10$ lowest energy states of BeH$_2$ (in the STO-3G basis) as function of bond distance. 

The $9$ excited states shown in Fig. \ref{fig:beh2_exc_fci_energies} are ordered in three degenerate energy levels.
The table below summarises the energy levels and their corresponding excited states at bond distances $r_{Be-H} = 1.5 \textup{\AA}$ and $r_{Be-H}=1.75 \textup{\AA}$:
\begin{table}[H]
\centering
\begin{tabular}{c}
  $r_{Be-H}=1.5 \text{\AA}$  \\
 \hline
 \begin{tabular}{c|c}
 exc. states & energy, Hartree \\
 \hline
 \{1,2\} & $E_I(1.5 \text{\AA}) \approx -15.3343$  \\
 \{3,4,5,6,7,8\} & $E_{II}(1.5 \text{\AA}) \approx-15.3331$  \\
 \{9\} & $E_{III}(1.5 \text{\AA}) \approx -15.3026$
 \end{tabular} \\
 \hline
  \hline
 $r_{Be-H}=1.75 \text{\AA}$ \\
  \hline

  \begin{tabular}{c|c}
  exc. states & energy, Hartree \\
 \hline
 \{1,2\} & $E_I(1.75 \text{\AA}) \approx-15.3131$  \\
 \{3,4,5\} & $E_{II}(1.75 \text{\AA})\approx-15.3059$  \\
  \{6,7,8,9\} & $E_{III}(1.75 \text{\AA})\approx-15.3056$  
 \end{tabular}\\
\end{tabular}
\caption{Energy levels for the $9$ lowest excited states of BeH$_2$ in the STO-3G basis. The ground state (not shown) is counted as the $0^{\text{th}}$.}
\end{table}
For $r_{Be-H}=1.5 \text{\AA}$ we can see that $E_{I}(1.5 \text{\AA})$ and $E_{II}(1.5 \text{\AA})$ are very close.
Upon expectation of the energy estimate of the e-QEB-ADAPT-VQE($\epsilon=10^{-8}$), for $r_{Be-H}=1.5 \text{\AA}$, it turns out that the algorithm actually finds $E_{II}(1.5 \text{\AA})$ with accuracy of $3 \times 10^{-8}$ Hartree.
Therefore, the problem is not a lack of accuracy, but converging to the wrong excited state.

Two possible reasons for this are: (1) the classical minimizers employed by the e-QEB-ADAPT-VQE get stuck in a local minimum corresponding to $E_{II}(1.5 \text{\AA})$, and/or (2) the e-QEB-ADAPT-VQE fails to construct an ansatz that can approximate the degenerate excited states corresponding to energy level $E_{I}(1.5 \text{\AA})$.
Since the e-QEB-ADAPT-VQE uses a combination of the BFGS and the Nelder-Mead optimization methods, where the Nelder-Mead is a direct-search method, which is unlikely to get stuck in local minimum, the first possible reason can be ruled out. Hence, we are left with the second one.

Upon explicit inspection of the degenerate states corresponding to energy level $E_I(1.5 \text{\AA})$, it is found that they have strong static correlations \cite{classical_methods}, where two dominant Slater determinants contributed equally to the wavefunction.
On the other hand for each of the degenerate states corresponding to energy level $E_{II}(1.5 \text{\AA})$ there is one dominant Slater determinant.
Hence, the first few ansatz-constructing-iterations of the e-QEB-ADAPT-VQE grow the ansatz with qubit excitation evolutions, which map between the initial reference state, $|\psi_0\rangle$, and the dominant Slater determinant of one of the states corresponding to $E_{II}(1.5 \text{\AA})$, because these qubit excitation evolutions decreases the estimate for the energy by the most. 
From that point on, the e-QEB-ADAPT-VQE continues to construct an ansatz that approximates a state corresponding to $E_{II}(1.5 \text{\AA})$, instead a state corresponding to $E_{I}(1.5 \text{\AA})$.

In the case of $r_{Be-H}=1.75 \text{\AA}$, the situation is similar. However, in this case the three energy levels $E_{I}(1.75 \text{\AA})$, $E_{II}(1.75 \text{\AA})$ and $E_{III}(1.75 \text{\AA})$ are close, and  the QEB-ADAPT-VQE converges to an excited state corresponding to $E_{III}(1.75 \text{\AA})$.
Again the states corresponding to $E_{I}(1.75 \text{\AA})$ and $E_{II}(1.75 \text{\AA})$ have two (not equally) dominant Slater determinants, whereas the states corresponding to $E_{III}(1.75 \text{\AA})$ have weaker static correlations, and just one dominant Slater determinant.

\onecolumngrid

\twocolumngrid

Overall the following conclusion can be made:
A generic feature of the e-QEB-ADAPT-VQE is to be more ``willing'' to construct ans\"atze for less statically correlated states (where one Slater determinant is dominant). This owes to its greedy strategy to grow its ansatz with the qubit excitation evolution that decrease the energy estimate by the most at each ansatz-growing iteration.
Hence, in the case when (1) the lowest energy eigenstates of an electronic Hamiltonian are ordered in two or more closely spaced (possibly degenerate) energy levels, and (2) the lowest energy level corresponds to eigenstates that have stronger static correlations than the eigenstates above them, the e-QEB-ADAPT-VQE might converge to one of the less statically correlated eigenstates, instead to one with the lowest energy.

\begin{figure}
\centering

\includegraphics[width=8cm]{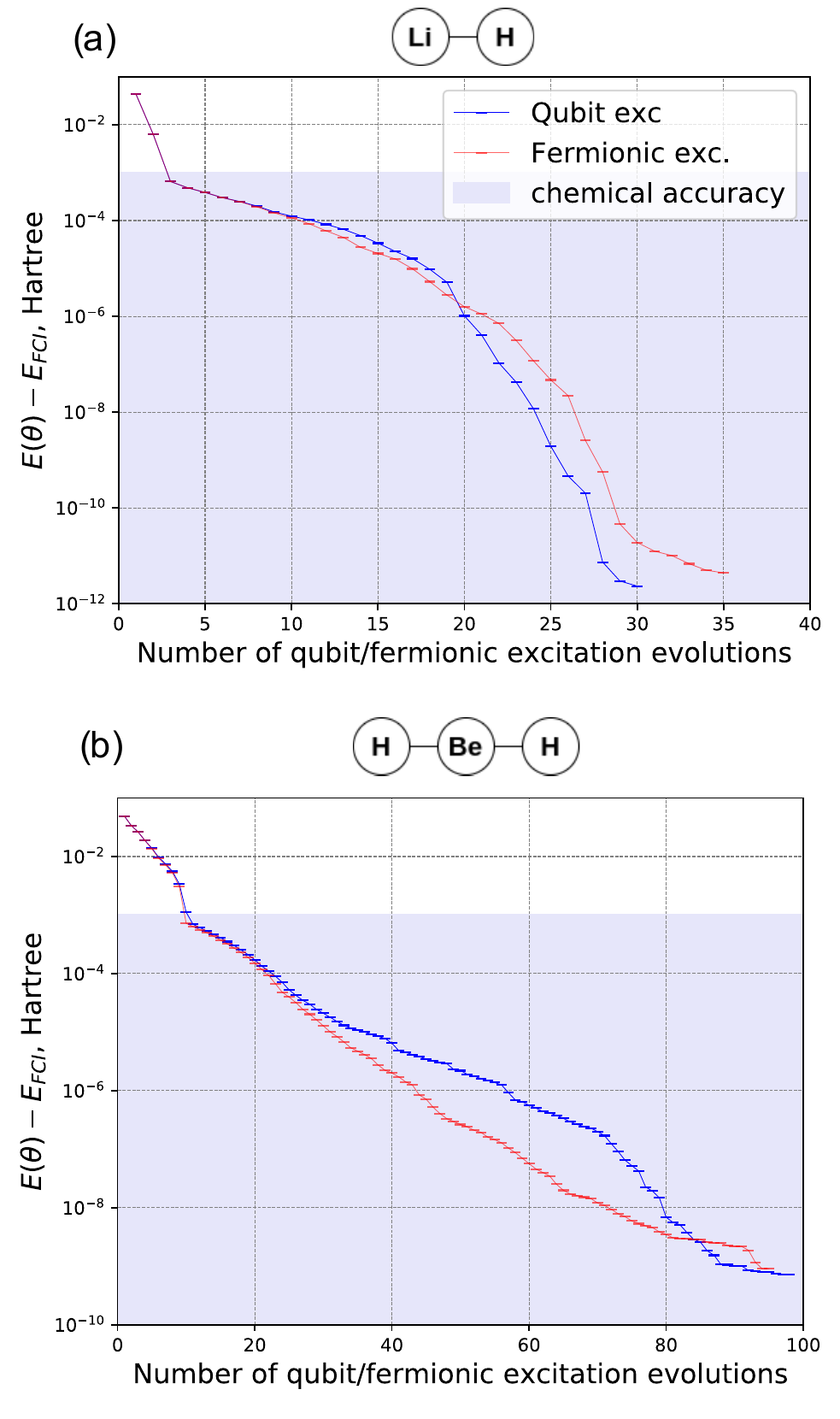}

\caption[Qubit evolutions versus fermionic evolutions for excited states]{Energy convergence plots for the first excited states of LiH and BeH$_2$ in the STO-3G basis at equilibrium bond distances of $r_{\text{Li-H}}=1.546 {\textup{\AA}}$ and $r_{\text{Be-H}}=1.316 {\textup{\AA}}$, respectively. The plots are obtained with the e-QEB-ADAPT-VQE for $n_{qe}=10$.}
\label{fig:exc_qe_vs_fe}
\end{figure}

\section{Qubit vs fermionic excitation evolutions}\label{sec:qe_vs_fe_excited states}

The use of qubit excitation evolutions by the QEB-ADAPT-VQE was motivated by the fact that they are implemented by simpler quantum circuits than fermionic excitation evolutions \cite{eff_circs, iqeb_vqe}, and the presumption that a qubit-excitation-based ansatz can approximate an electronic wavefunction as well as a fermionic-excitation-based ansatz.
In Ref. \cite{iqeb_vqe} we compared the two types of ans\"atze in approximating the ground states for small molecules, and found that the ans\"atze perform similarly, in terms of accuracy. 
In this section, we generalise this comparison for (more strongly correlated) excited states.

Figure \ref{fig:exc_qe_vs_fe} shows energy convergence plots for the first excited states of LiH and BeH$_2$, at bond distances of $r_{\text{Li-H}}=1.546 {\textup{\AA}}$ and $r_{\text{Be-H}}=1.316 {\textup{\AA}}$, respectively.
The blue plots are obtained with the e-QEB-ADAPT-VQE for $n=10$ (the same plots as in Fig. \ref{sec:exc_iqeb_vs_iqeb}), and the red plots are obtained with the e-QEB-ADAPT-VQE ansatz-constructing routine, for $n=10$, using a pool of single and double fermionic (instead of qubit) excitation evolutions.
We can observe close similarity between the two types of ans\"atze.
In the case of LiH (Fig. \ref{fig:exc_qe_vs_fe}a) the qubit-excitation-based ansatz is slightly more accurate per number of ansatz elements, while in the case of BeH$_2$ (Fig. \ref{fig:exc_qe_vs_fe}b) the fermionic-excitation-based ansatz is a bit more accurate instead.
These differences are relatively small, so we derive the same conclusion as in Refs. \cite{iqeb_vqe, UCCSD_qubit}, that qubit and fermionic excitation evolutions can approximate electronic wavefunctions, comparably well.

\section{Conclusion}

In this paper we proposed a modified version of the QEB-ADAPT-VQE algorithm, the e-QEB-ADAPT-VQE, designed to simulate low-lying excited molecular states.
 
The QEB-ADAPT-VQE relies on an initial reference state, which has a large overlap with the target ground state in order to construct an efficient ansatz.
However, choosing an initial reference state that has a significant overlap with a target excited state is not as straightforward as for a target ground state. Therefore, the e-QEB-ADAPT-VQE is designed to be less dependent on the choice of an initial reference state. 
This is achieved by an ansatz growing strategy that does not rely on energy-gradient evaluations, like the QEB-ADAPT-VQE, but instead on individual energy-reduction evaluations.
The modified ansatz growing strategy comes at a cost of up to a constant factor of more quantum computer measurements, as compared to the original QEB-ADAPT-VQE.

We benchmarked the performance of the e-QEB-ADAPT-VQE with classical numerical simulations, by constructing ans\"atze for the first excited states of LiH and BeH$_2$. 
We found that the e-QEB-ADAPT-VQE can construct highly accurate ans\"atze that are implemented by multiple times shallower circuits, which require multiple times fewer $CNOT$s, than fixed UCC ans\"atze, like the UCCSD and the GUCCSD.
Therefore, the e-QEB-ADAPT-VQE is especially suitable for NISQ computers, for which the number of $2$-qubits entangling gates, such as $CNOT$s, that can be applied reliably is the current bottleneck \cite{NISQ}.

We also found that the e-QEB-ADAPT-VQE might fail to find excited-state energies in order of increasing energy, if the excited states are ordered in closely spaced degenerate energy levels, and the lower lying excited states are more statically correlated than the states above them.
This problem derives from the fact that the e-QEB-ADAPT-VQE pursues the greedy strategy to achieve a lowest estimate for the energy at each ansatz-growing iteration. Hence, the algorithm runs the risk of going along the wrong path and converge to a state, which is not necessarily the one with the lowest energy.
In fact, this problem will be present in any iterative-VQE protocol that relies on such a greedy strategy, including the QEB-ADAPT-VQE, the ADAPT-VQE \cite{adapt_vqe} and the qubit-ADAPT-VQE \cite{qubit_adapt_vqe}.
Nevertheless, this problem is unlikely to be significant, because in practice one would be interested in finding the whole spectrum of low lying excited state energies, so the order in which they are found is not necessarily important.

Lastly, we used the e-QEB-ADAPT-VQE ansatz-constructing routine to compare qubit-excitation-based and fermionic-excitation-based ans\"atze in approximating excited states. The two types of ans\"atze were found to perform similarly, achieving a particular accuracy with approximately the same number of qubit/fermionic excitation evolutions.
These observations generalize previous results \cite{iqeb_vqe, UCCSD_qubit} about the equivalence of qubit and fermionic excitation evolutions in approximating electronic wavefunctions.

\acknowledgements
The authors wish to thank Dr N. Mertig for useful discussions.
 Y.S.Y. acknowledges financial support from the EPSRC and Hitachi via CASE studentships RG97399.

\bibliographystyle{apsrev4-1}
\bibliography{references}

\begin{thebibliography}{37}%
\makeatletter
\providecommand \@ifxundefined [1]{%
 \@ifx{#1\undefined}
}%
\providecommand \@ifnum [1]{%
 \ifnum #1\expandafter \@firstoftwo
 \else \expandafter \@secondoftwo
 \fi
}%
\providecommand \@ifx [1]{%
 \ifx #1\expandafter \@firstoftwo
 \else \expandafter \@secondoftwo
 \fi
}%
\providecommand \natexlab [1]{#1}%
\providecommand \enquote  [1]{``#1''}%
\providecommand \bibnamefont  [1]{#1}%
\providecommand \bibfnamefont [1]{#1}%
\providecommand \citenamefont [1]{#1}%
\providecommand \href@noop [0]{\@secondoftwo}%
\providecommand \href [0]{\begingroup \@sanitize@url \@href}%
\providecommand \@href[1]{\@@startlink{#1}\@@href}%
\providecommand \@@href[1]{\endgroup#1\@@endlink}%
\providecommand \@sanitize@url [0]{\catcode `\\12\catcode `\$12\catcode
  `\&12\catcode `\#12\catcode `\^12\catcode `\_12\catcode `\%12\relax}%
\providecommand \@@startlink[1]{}%
\providecommand \@@endlink[0]{}%
\providecommand \url  [0]{\begingroup\@sanitize@url \@url }%
\providecommand \@url [1]{\endgroup\@href {#1}{\urlprefix }}%
\providecommand \urlprefix  [0]{URL }%
\providecommand \Eprint [0]{\href }%
\providecommand \doibase [0]{http://dx.doi.org/}%
\providecommand \selectlanguage [0]{\@gobble}%
\providecommand \bibinfo  [0]{\@secondoftwo}%
\providecommand \bibfield  [0]{\@secondoftwo}%
\providecommand \translation [1]{[#1]}%
\providecommand \BibitemOpen [0]{}%
\providecommand \bibitemStop [0]{}%
\providecommand \bibitemNoStop [0]{.\EOS\space}%
\providecommand \EOS [0]{\spacefactor3000\relax}%
\providecommand \BibitemShut  [1]{\csname bibitem#1\endcsname}%
\let\auto@bib@innerbib\@empty
\bibitem [{\citenamefont {McArdle}\ \emph {et~al.}(2020)\citenamefont
  {McArdle}, \citenamefont {Endo}, \citenamefont {Aspuru-Guzik}, \citenamefont
  {Benjamin},\ and\ \citenamefont {Yuan}}]{vqe_general}%
  \BibitemOpen
  \bibfield  {author} {\bibinfo {author} {\bibfnamefont {S.}~\bibnamefont
  {McArdle}}, \bibinfo {author} {\bibfnamefont {S.}~\bibnamefont {Endo}},
  \bibinfo {author} {\bibfnamefont {A.}~\bibnamefont {Aspuru-Guzik}}, \bibinfo
  {author} {\bibfnamefont {S.~C.}\ \bibnamefont {Benjamin}}, \ and\ \bibinfo
  {author} {\bibfnamefont {X.}~\bibnamefont {Yuan}},\ }\href {\doibase
  10.1103/RevModPhys.92.015003} {\bibfield  {journal} {\bibinfo  {journal}
  {Rev. Mod. Phys.}\ }\textbf {\bibinfo {volume} {92}},\ \bibinfo {pages}
  {015003} (\bibinfo {year} {2020})}\BibitemShut {NoStop}%
\bibitem [{\citenamefont {McClean}\ \emph {et~al.}(2016)\citenamefont
  {McClean}, \citenamefont {Romero}, \citenamefont {Babbush},\ and\
  \citenamefont {Aspuru-Guzik}}]{vqe_general_2}%
  \BibitemOpen
  \bibfield  {author} {\bibinfo {author} {\bibfnamefont {J.~R.}\ \bibnamefont
  {McClean}}, \bibinfo {author} {\bibfnamefont {J.}~\bibnamefont {Romero}},
  \bibinfo {author} {\bibfnamefont {R.}~\bibnamefont {Babbush}}, \ and\
  \bibinfo {author} {\bibfnamefont {A.}~\bibnamefont {Aspuru-Guzik}},\ }\href
  {\doibase https://doi.org/10.1088/1367-2630/18/2/023023} {\bibfield
  {journal} {\bibinfo  {journal} {New Journal of Physics}\ }\textbf {\bibinfo
  {volume} {18}},\ \bibinfo {pages} {023023} (\bibinfo {year}
  {2016})}\BibitemShut {NoStop}%
\bibitem [{\citenamefont {O'Malley}\ \emph {et~al.}(2016)\citenamefont
  {O'Malley}, \citenamefont {Babbush}, \citenamefont {Kivlichan}, \citenamefont
  {Romero}, \citenamefont {McClean}, \citenamefont {Barends}, \citenamefont
  {Kelly}, \citenamefont {Roushan}, \citenamefont {Tranter}, \citenamefont
  {Ding}, \citenamefont {Campbell}, \citenamefont {Chen}, \citenamefont {Chen},
  \citenamefont {Chiaro}, \citenamefont {Dunsworth}, \citenamefont {Fowler},
  \citenamefont {Jeffrey}, \citenamefont {Lucero}, \citenamefont {Megrant},
  \citenamefont {Mutus}, \citenamefont {Neeley}, \citenamefont {Neill},
  \citenamefont {Quintana}, \citenamefont {Sank}, \citenamefont {Vainsencher},
  \citenamefont {Wenner}, \citenamefont {White}, \citenamefont {Coveney},
  \citenamefont {Love}, \citenamefont {Neven}, \citenamefont {Aspuru-Guzik},\
  and\ \citenamefont {Martinis}}]{vqe_2}%
  \BibitemOpen
  \bibfield  {author} {\bibinfo {author} {\bibfnamefont {P.~J.~J.}\
  \bibnamefont {O'Malley}}, \bibinfo {author} {\bibfnamefont {R.}~\bibnamefont
  {Babbush}}, \bibinfo {author} {\bibfnamefont {I.~D.}\ \bibnamefont
  {Kivlichan}}, \bibinfo {author} {\bibfnamefont {J.}~\bibnamefont {Romero}},
  \bibinfo {author} {\bibfnamefont {J.~R.}\ \bibnamefont {McClean}}, \bibinfo
  {author} {\bibfnamefont {R.}~\bibnamefont {Barends}}, \bibinfo {author}
  {\bibfnamefont {J.}~\bibnamefont {Kelly}}, \bibinfo {author} {\bibfnamefont
  {P.}~\bibnamefont {Roushan}}, \bibinfo {author} {\bibfnamefont
  {A.}~\bibnamefont {Tranter}}, \bibinfo {author} {\bibfnamefont
  {N.}~\bibnamefont {Ding}}, \bibinfo {author} {\bibfnamefont {B.}~\bibnamefont
  {Campbell}}, \bibinfo {author} {\bibfnamefont {Y.}~\bibnamefont {Chen}},
  \bibinfo {author} {\bibfnamefont {Z.}~\bibnamefont {Chen}}, \bibinfo {author}
  {\bibfnamefont {B.}~\bibnamefont {Chiaro}}, \bibinfo {author} {\bibfnamefont
  {A.}~\bibnamefont {Dunsworth}}, \bibinfo {author} {\bibfnamefont {A.~G.}\
  \bibnamefont {Fowler}}, \bibinfo {author} {\bibfnamefont {E.}~\bibnamefont
  {Jeffrey}}, \bibinfo {author} {\bibfnamefont {E.}~\bibnamefont {Lucero}},
  \bibinfo {author} {\bibfnamefont {A.}~\bibnamefont {Megrant}}, \bibinfo
  {author} {\bibfnamefont {J.~Y.}\ \bibnamefont {Mutus}}, \bibinfo {author}
  {\bibfnamefont {M.}~\bibnamefont {Neeley}}, \bibinfo {author} {\bibfnamefont
  {C.}~\bibnamefont {Neill}}, \bibinfo {author} {\bibfnamefont
  {C.}~\bibnamefont {Quintana}}, \bibinfo {author} {\bibfnamefont
  {D.}~\bibnamefont {Sank}}, \bibinfo {author} {\bibfnamefont {A.}~\bibnamefont
  {Vainsencher}}, \bibinfo {author} {\bibfnamefont {J.}~\bibnamefont {Wenner}},
  \bibinfo {author} {\bibfnamefont {T.~C.}\ \bibnamefont {White}}, \bibinfo
  {author} {\bibfnamefont {P.~V.}\ \bibnamefont {Coveney}}, \bibinfo {author}
  {\bibfnamefont {P.~J.}\ \bibnamefont {Love}}, \bibinfo {author}
  {\bibfnamefont {H.}~\bibnamefont {Neven}}, \bibinfo {author} {\bibfnamefont
  {A.}~\bibnamefont {Aspuru-Guzik}}, \ and\ \bibinfo {author} {\bibfnamefont
  {J.~M.}\ \bibnamefont {Martinis}},\ }\href {\doibase
  10.1103/PhysRevX.6.031007} {\bibfield  {journal} {\bibinfo  {journal} {Phys.
  Rev. X}\ }\textbf {\bibinfo {volume} {6}},\ \bibinfo {pages} {031007}
  (\bibinfo {year} {2016})}\BibitemShut {NoStop}%
\bibitem [{\citenamefont {Wang}\ \emph {et~al.}(2019)\citenamefont {Wang},
  \citenamefont {Higgott},\ and\ \citenamefont {Brierley}}]{VQE_accelerated}%
  \BibitemOpen
  \bibfield  {author} {\bibinfo {author} {\bibfnamefont {D.}~\bibnamefont
  {Wang}}, \bibinfo {author} {\bibfnamefont {O.}~\bibnamefont {Higgott}}, \
  and\ \bibinfo {author} {\bibfnamefont {S.}~\bibnamefont {Brierley}},\ }\href
  {\doibase 10.1103/PhysRevLett.122.140504} {\bibfield  {journal} {\bibinfo
  {journal} {Phys. Rev. Lett.}\ }\textbf {\bibinfo {volume} {122}},\ \bibinfo
  {pages} {140504} (\bibinfo {year} {2019})}\BibitemShut {NoStop}%
\bibitem [{\citenamefont {Arute}\ \emph {et~al.}(2020)\citenamefont {Arute},
  \citenamefont {Arya}, \citenamefont {Babbush}, \citenamefont {Bacon},
  \citenamefont {Bardin}, \citenamefont {Barends}, \citenamefont {Boixo},
  \citenamefont {Broughton}, \citenamefont {Buckley}, \citenamefont {Buell}
  \emph {et~al.}}]{vqe_google_hf}%
  \BibitemOpen
  \bibfield  {author} {\bibinfo {author} {\bibfnamefont {F.}~\bibnamefont
  {Arute}}, \bibinfo {author} {\bibfnamefont {K.}~\bibnamefont {Arya}},
  \bibinfo {author} {\bibfnamefont {R.}~\bibnamefont {Babbush}}, \bibinfo
  {author} {\bibfnamefont {D.}~\bibnamefont {Bacon}}, \bibinfo {author}
  {\bibfnamefont {J.~C.}\ \bibnamefont {Bardin}}, \bibinfo {author}
  {\bibfnamefont {R.}~\bibnamefont {Barends}}, \bibinfo {author} {\bibfnamefont
  {S.}~\bibnamefont {Boixo}}, \bibinfo {author} {\bibfnamefont
  {M.}~\bibnamefont {Broughton}}, \bibinfo {author} {\bibfnamefont {B.~B.}\
  \bibnamefont {Buckley}}, \bibinfo {author} {\bibfnamefont {D.~A.}\
  \bibnamefont {Buell}},  \emph {et~al.},\ }\href {\doibase
  10.1126/science.abb9811} {\bibfield  {journal} {\bibinfo  {journal} {arXiv
  preprint arXiv:2004.04174}\ } (\bibinfo {year} {2020}),\
  10.1126/science.abb9811}\BibitemShut {NoStop}%
\bibitem [{\citenamefont {Preskill}(2018)}]{NISQ}%
  \BibitemOpen
  \bibfield  {author} {\bibinfo {author} {\bibfnamefont {J.}~\bibnamefont
  {Preskill}},\ }\href {\doibase https://doi.org/10.22331/q-2018-08-06-79}
  {\bibfield  {journal} {\bibinfo  {journal} {Quantum}\ }\textbf {\bibinfo
  {volume} {2}},\ \bibinfo {pages} {79} (\bibinfo {year} {2018})}\BibitemShut
  {NoStop}%
\bibitem [{\citenamefont {Arute}\ \emph {et~al.}(2019)\citenamefont {Arute},
  \citenamefont {Arya}, \citenamefont {Babbush}, \citenamefont {Bacon},
  \citenamefont {Bardin}, \citenamefont {Barends}, \citenamefont {Biswas},
  \citenamefont {Boixo}, \citenamefont {Brandao}, \citenamefont {Buell} \emph
  {et~al.}}]{google_supreme}%
  \BibitemOpen
  \bibfield  {author} {\bibinfo {author} {\bibfnamefont {F.}~\bibnamefont
  {Arute}}, \bibinfo {author} {\bibfnamefont {K.}~\bibnamefont {Arya}},
  \bibinfo {author} {\bibfnamefont {R.}~\bibnamefont {Babbush}}, \bibinfo
  {author} {\bibfnamefont {D.}~\bibnamefont {Bacon}}, \bibinfo {author}
  {\bibfnamefont {J.~C.}\ \bibnamefont {Bardin}}, \bibinfo {author}
  {\bibfnamefont {R.}~\bibnamefont {Barends}}, \bibinfo {author} {\bibfnamefont
  {R.}~\bibnamefont {Biswas}}, \bibinfo {author} {\bibfnamefont
  {S.}~\bibnamefont {Boixo}}, \bibinfo {author} {\bibfnamefont {F.~G.}\
  \bibnamefont {Brandao}}, \bibinfo {author} {\bibfnamefont {D.~A.}\
  \bibnamefont {Buell}},  \emph {et~al.},\ }\href {\doibase
  https://doi.org/10.1038/s41586-019-1666-5} {\bibfield  {journal} {\bibinfo
  {journal} {Nature}\ }\textbf {\bibinfo {volume} {574}},\ \bibinfo {pages}
  {505} (\bibinfo {year} {2019})}\BibitemShut {NoStop}%
\bibitem [{\citenamefont {Elfving}\ \emph {et~al.}(2020)\citenamefont
  {Elfving}, \citenamefont {Broer}, \citenamefont {Webber}, \citenamefont
  {Gavartin}, \citenamefont {Halls}, \citenamefont {Lorton},\ and\
  \citenamefont {Bochevarov}}]{vqe_nisq}%
  \BibitemOpen
  \bibfield  {author} {\bibinfo {author} {\bibfnamefont {V.~E.}\ \bibnamefont
  {Elfving}}, \bibinfo {author} {\bibfnamefont {B.~W.}\ \bibnamefont {Broer}},
  \bibinfo {author} {\bibfnamefont {M.}~\bibnamefont {Webber}}, \bibinfo
  {author} {\bibfnamefont {J.}~\bibnamefont {Gavartin}}, \bibinfo {author}
  {\bibfnamefont {M.~D.}\ \bibnamefont {Halls}}, \bibinfo {author}
  {\bibfnamefont {K.~P.}\ \bibnamefont {Lorton}}, \ and\ \bibinfo {author}
  {\bibfnamefont {A.}~\bibnamefont {Bochevarov}},\ }\href
  {https://arxiv.org/abs/2009.12472} {\  (\bibinfo {year} {2020})},\ \Eprint
  {http://arxiv.org/abs/2009.12472} {arXiv:2009.12472} \BibitemShut {NoStop}%
\bibitem [{\citenamefont {Helgaker}\ \emph {et~al.}(2014)\citenamefont
  {Helgaker}, \citenamefont {Jorgensen},\ and\ \citenamefont {Olsen}}]{CCSD}%
  \BibitemOpen
  \bibfield  {author} {\bibinfo {author} {\bibfnamefont {T.}~\bibnamefont
  {Helgaker}}, \bibinfo {author} {\bibfnamefont {P.}~\bibnamefont {Jorgensen}},
  \ and\ \bibinfo {author} {\bibfnamefont {J.}~\bibnamefont {Olsen}},\ }\href
  {https://onlinelibrary.wiley.com/doi/pdf/10.1002/9781119019572.fmatter}
  {\emph {\bibinfo {title} {Molecular electronic-structure theory}}}\ (\bibinfo
   {publisher} {John Wiley \& Sons},\ \bibinfo {year} {2014})\BibitemShut
  {NoStop}%
\bibitem [{\citenamefont {Dorner}\ \emph {et~al.}(2009)\citenamefont {Dorner},
  \citenamefont {Demkowicz-Dobrzanski}, \citenamefont {Smith}, \citenamefont
  {Lundeen}, \citenamefont {Wasilewski}, \citenamefont {Banaszek},\ and\
  \citenamefont {Walmsley}}]{QPE}%
  \BibitemOpen
  \bibfield  {author} {\bibinfo {author} {\bibfnamefont {U.}~\bibnamefont
  {Dorner}}, \bibinfo {author} {\bibfnamefont {R.}~\bibnamefont
  {Demkowicz-Dobrzanski}}, \bibinfo {author} {\bibfnamefont {B.~J.}\
  \bibnamefont {Smith}}, \bibinfo {author} {\bibfnamefont {J.~S.}\ \bibnamefont
  {Lundeen}}, \bibinfo {author} {\bibfnamefont {W.}~\bibnamefont {Wasilewski}},
  \bibinfo {author} {\bibfnamefont {K.}~\bibnamefont {Banaszek}}, \ and\
  \bibinfo {author} {\bibfnamefont {I.~A.}\ \bibnamefont {Walmsley}},\ }\href
  {\doibase 10.1103/PhysRevLett.102.040403} {\bibfield  {journal} {\bibinfo
  {journal} {Phys. Rev. Lett.}\ }\textbf {\bibinfo {volume} {102}},\ \bibinfo
  {pages} {040403} (\bibinfo {year} {2009})}\BibitemShut {NoStop}%
\bibitem [{\citenamefont {Hempel}\ \emph {et~al.}(2018)\citenamefont {Hempel},
  \citenamefont {Maier}, \citenamefont {Romero}, \citenamefont {McClean},
  \citenamefont {Monz}, \citenamefont {Shen}, \citenamefont {Jurcevic},
  \citenamefont {Lanyon}, \citenamefont {Love}, \citenamefont {Babbush},
  \citenamefont {Aspuru-Guzik}, \citenamefont {Blatt},\ and\ \citenamefont
  {Roos}}]{UCCSD}%
  \BibitemOpen
  \bibfield  {author} {\bibinfo {author} {\bibfnamefont {C.}~\bibnamefont
  {Hempel}}, \bibinfo {author} {\bibfnamefont {C.}~\bibnamefont {Maier}},
  \bibinfo {author} {\bibfnamefont {J.}~\bibnamefont {Romero}}, \bibinfo
  {author} {\bibfnamefont {J.}~\bibnamefont {McClean}}, \bibinfo {author}
  {\bibfnamefont {T.}~\bibnamefont {Monz}}, \bibinfo {author} {\bibfnamefont
  {H.}~\bibnamefont {Shen}}, \bibinfo {author} {\bibfnamefont {P.}~\bibnamefont
  {Jurcevic}}, \bibinfo {author} {\bibfnamefont {B.~P.}\ \bibnamefont
  {Lanyon}}, \bibinfo {author} {\bibfnamefont {P.}~\bibnamefont {Love}},
  \bibinfo {author} {\bibfnamefont {R.}~\bibnamefont {Babbush}}, \bibinfo
  {author} {\bibfnamefont {A.}~\bibnamefont {Aspuru-Guzik}}, \bibinfo {author}
  {\bibfnamefont {R.}~\bibnamefont {Blatt}}, \ and\ \bibinfo {author}
  {\bibfnamefont {C.~F.}\ \bibnamefont {Roos}},\ }\href {\doibase
  10.1103/PhysRevX.8.031022} {\bibfield  {journal} {\bibinfo  {journal} {Phys.
  Rev. X}\ }\textbf {\bibinfo {volume} {8}},\ \bibinfo {pages} {031022}
  (\bibinfo {year} {2018})}\BibitemShut {NoStop}%
\bibitem [{\citenamefont {Romero}\ \emph {et~al.}(2018)\citenamefont {Romero},
  \citenamefont {Babbush}, \citenamefont {McClean}, \citenamefont {Hempel},
  \citenamefont {Love},\ and\ \citenamefont {Aspuru-Guzik}}]{UCCSD_2}%
  \BibitemOpen
  \bibfield  {author} {\bibinfo {author} {\bibfnamefont {J.}~\bibnamefont
  {Romero}}, \bibinfo {author} {\bibfnamefont {R.}~\bibnamefont {Babbush}},
  \bibinfo {author} {\bibfnamefont {J.~R.}\ \bibnamefont {McClean}}, \bibinfo
  {author} {\bibfnamefont {C.}~\bibnamefont {Hempel}}, \bibinfo {author}
  {\bibfnamefont {P.~J.}\ \bibnamefont {Love}}, \ and\ \bibinfo {author}
  {\bibfnamefont {A.}~\bibnamefont {Aspuru-Guzik}},\ }\href {\doibase
  https://doi.org/10.1088/2058-9565/aad3e4} {\bibfield  {journal} {\bibinfo
  {journal} {Quantum Science and Technology}\ }\textbf {\bibinfo {volume}
  {4}},\ \bibinfo {pages} {014008} (\bibinfo {year} {2018})}\BibitemShut
  {NoStop}%
\bibitem [{\citenamefont {Lee}\ \emph {et~al.}(2018)\citenamefont {Lee},
  \citenamefont {Huggins}, \citenamefont {Head-Gordon},\ and\ \citenamefont
  {Whaley}}]{k-upCCGSD}%
  \BibitemOpen
  \bibfield  {author} {\bibinfo {author} {\bibfnamefont {J.}~\bibnamefont
  {Lee}}, \bibinfo {author} {\bibfnamefont {W.~J.}\ \bibnamefont {Huggins}},
  \bibinfo {author} {\bibfnamefont {M.}~\bibnamefont {Head-Gordon}}, \ and\
  \bibinfo {author} {\bibfnamefont {K.~B.}\ \bibnamefont {Whaley}},\ }\href
  {\doibase https://doi.org/10.1021/acs.jctc.8b01004} {\bibfield  {journal}
  {\bibinfo  {journal} {Journal of chemical theory and computation}\ }\textbf
  {\bibinfo {volume} {15}},\ \bibinfo {pages} {311} (\bibinfo {year}
  {2018})}\BibitemShut {NoStop}%
\bibitem [{\citenamefont {Dallaire-Demers}\ \emph {et~al.}(2019)\citenamefont
  {Dallaire-Demers}, \citenamefont {Romero}, \citenamefont {Veis},
  \citenamefont {Sim},\ and\ \citenamefont {Aspuru-Guzik}}]{UCCSD_bogoliubov}%
  \BibitemOpen
  \bibfield  {author} {\bibinfo {author} {\bibfnamefont {P.-L.}\ \bibnamefont
  {Dallaire-Demers}}, \bibinfo {author} {\bibfnamefont {J.}~\bibnamefont
  {Romero}}, \bibinfo {author} {\bibfnamefont {L.}~\bibnamefont {Veis}},
  \bibinfo {author} {\bibfnamefont {S.}~\bibnamefont {Sim}}, \ and\ \bibinfo
  {author} {\bibfnamefont {A.}~\bibnamefont {Aspuru-Guzik}},\ }\href {\doibase
  https://doi.org/10.1088/2058-9565/ab3951} {\bibfield  {journal} {\bibinfo
  {journal} {Quantum Science and Technology}\ }\textbf {\bibinfo {volume}
  {4}},\ \bibinfo {pages} {045005} (\bibinfo {year} {2019})}\BibitemShut
  {NoStop}%
\bibitem [{\citenamefont {Sokolov}\ \emph {et~al.}(2020)\citenamefont
  {Sokolov}, \citenamefont {Barkoutsos}, \citenamefont {Ollitrault},
  \citenamefont {Greenberg}, \citenamefont {Rice}, \citenamefont {Pistoia},\
  and\ \citenamefont {Tavernelli}}]{optimized_uccsd}%
  \BibitemOpen
  \bibfield  {author} {\bibinfo {author} {\bibfnamefont {I.~O.}\ \bibnamefont
  {Sokolov}}, \bibinfo {author} {\bibfnamefont {P.~K.}\ \bibnamefont
  {Barkoutsos}}, \bibinfo {author} {\bibfnamefont {P.~J.}\ \bibnamefont
  {Ollitrault}}, \bibinfo {author} {\bibfnamefont {D.}~\bibnamefont
  {Greenberg}}, \bibinfo {author} {\bibfnamefont {J.}~\bibnamefont {Rice}},
  \bibinfo {author} {\bibfnamefont {M.}~\bibnamefont {Pistoia}}, \ and\
  \bibinfo {author} {\bibfnamefont {I.}~\bibnamefont {Tavernelli}},\ }\href
  {\doibase 10.1063/1.5141835} {\bibfield  {journal} {\bibinfo  {journal} {The
  Journal of Chemical Physics}\ }\textbf {\bibinfo {volume} {152}},\ \bibinfo
  {pages} {124107} (\bibinfo {year} {2020})}\BibitemShut {NoStop}%
\bibitem [{\citenamefont {Peruzzo}\ \emph {et~al.}(2014)\citenamefont
  {Peruzzo}, \citenamefont {McClean}, \citenamefont {Shadbolt}, \citenamefont
  {Yung}, \citenamefont {Zhou}, \citenamefont {Love}, \citenamefont
  {Aspuru-Guzik},\ and\ \citenamefont {O’brien}}]{vqe_first_uccsd}%
  \BibitemOpen
  \bibfield  {author} {\bibinfo {author} {\bibfnamefont {A.}~\bibnamefont
  {Peruzzo}}, \bibinfo {author} {\bibfnamefont {J.}~\bibnamefont {McClean}},
  \bibinfo {author} {\bibfnamefont {P.}~\bibnamefont {Shadbolt}}, \bibinfo
  {author} {\bibfnamefont {M.-H.}\ \bibnamefont {Yung}}, \bibinfo {author}
  {\bibfnamefont {X.-Q.}\ \bibnamefont {Zhou}}, \bibinfo {author}
  {\bibfnamefont {P.~J.}\ \bibnamefont {Love}}, \bibinfo {author}
  {\bibfnamefont {A.}~\bibnamefont {Aspuru-Guzik}}, \ and\ \bibinfo {author}
  {\bibfnamefont {J.~L.}\ \bibnamefont {O’brien}},\ }\href {\doibase
  https://doi.org/10.1038/ncomms5213} {\bibfield  {journal} {\bibinfo
  {journal} {Nature communications}\ }\textbf {\bibinfo {volume} {5}},\
  \bibinfo {pages} {4213} (\bibinfo {year} {2014})}\BibitemShut {NoStop}%
\bibitem [{\citenamefont {Grimsley}\ \emph {et~al.}(2019)\citenamefont
  {Grimsley}, \citenamefont {Economou}, \citenamefont {Barnes},\ and\
  \citenamefont {Mayhall}}]{adapt_vqe}%
  \BibitemOpen
  \bibfield  {author} {\bibinfo {author} {\bibfnamefont {H.~R.}\ \bibnamefont
  {Grimsley}}, \bibinfo {author} {\bibfnamefont {S.~E.}\ \bibnamefont
  {Economou}}, \bibinfo {author} {\bibfnamefont {E.}~\bibnamefont {Barnes}}, \
  and\ \bibinfo {author} {\bibfnamefont {N.~J.}\ \bibnamefont {Mayhall}},\
  }\href {\doibase https://doi.org/10.1038/s41467-019-10988-2} {\bibfield
  {journal} {\bibinfo  {journal} {Nature communications}\ }\textbf {\bibinfo
  {volume} {10}},\ \bibinfo {pages} {1} (\bibinfo {year} {2019})}\BibitemShut
  {NoStop}%
\bibitem [{\citenamefont {Tang}\ \emph {et~al.}(2019)\citenamefont {Tang},
  \citenamefont {Barnes}, \citenamefont {Grimsley}, \citenamefont {Mayhall},\
  and\ \citenamefont {Economou}}]{qubit_adapt_vqe}%
  \BibitemOpen
  \bibfield  {author} {\bibinfo {author} {\bibfnamefont {H.~L.}\ \bibnamefont
  {Tang}}, \bibinfo {author} {\bibfnamefont {E.}~\bibnamefont {Barnes}},
  \bibinfo {author} {\bibfnamefont {H.~R.}\ \bibnamefont {Grimsley}}, \bibinfo
  {author} {\bibfnamefont {N.~J.}\ \bibnamefont {Mayhall}}, \ and\ \bibinfo
  {author} {\bibfnamefont {S.~E.}\ \bibnamefont {Economou}},\ }\href
  {https://arxiv.org/pdf/1911.10205.pdf} {\bibfield  {journal} {\bibinfo
  {journal} {arXiv preprint arXiv:1911.10205}\ } (\bibinfo {year}
  {2019})}\BibitemShut {NoStop}%
\bibitem [{\citenamefont {Sim}\ \emph {et~al.}(2020)\citenamefont {Sim},
  \citenamefont {Romero}, \citenamefont {Gonthier},\ and\ \citenamefont
  {Kunitsa}}]{pruning_vqe}%
  \BibitemOpen
  \bibfield  {author} {\bibinfo {author} {\bibfnamefont {S.}~\bibnamefont
  {Sim}}, \bibinfo {author} {\bibfnamefont {J.}~\bibnamefont {Romero}},
  \bibinfo {author} {\bibfnamefont {J.~F.}\ \bibnamefont {Gonthier}}, \ and\
  \bibinfo {author} {\bibfnamefont {A.~A.}\ \bibnamefont {Kunitsa}},\ }\href
  {https://arxiv.org/abs/2010.00629} {\bibfield  {journal} {\bibinfo  {journal}
  {arXiv preprint arXiv:2010.00629}\ } (\bibinfo {year} {2020})}\BibitemShut
  {NoStop}%
\bibitem [{\citenamefont {Daniel~Claudino}\ and\ \citenamefont
  {Humble}(2020)}]{benchmark_adapt_vqe}%
  \BibitemOpen
  \bibfield  {author} {\bibinfo {author} {\bibfnamefont {A.~J.~M.}\
  \bibnamefont {Daniel~Claudino}, \bibfnamefont {Jerimiah~Wright}}\ and\
  \bibinfo {author} {\bibfnamefont {T.~S.}\ \bibnamefont {Humble}},\ }\href
  {https://arxiv.org/abs/2011.01279} {\bibfield  {journal} {\bibinfo  {journal}
  {arXiv preprint arXiv:2011.01279}\ } (\bibinfo {year} {2020})}\BibitemShut
  {NoStop}%
\bibitem [{\citenamefont {Yordanov}\ \emph {et~al.}(2021)\citenamefont
  {Yordanov}, \citenamefont {Armaos}, \citenamefont {Barnes},\ and\
  \citenamefont {Arvidsson-Shukur}}]{iqeb_vqe}%
  \BibitemOpen
  \bibfield  {author} {\bibinfo {author} {\bibfnamefont {Y.~S.}\ \bibnamefont
  {Yordanov}}, \bibinfo {author} {\bibfnamefont {V.}~\bibnamefont {Armaos}},
  \bibinfo {author} {\bibfnamefont {C.~H.}\ \bibnamefont {Barnes}}, \ and\
  \bibinfo {author} {\bibfnamefont {D.~R.}\ \bibnamefont {Arvidsson-Shukur}},\
  }\href {\doibase https://doi.org/10.1038/s42005-021-00730-0} {\bibfield
  {journal} {\bibinfo  {journal} {Commun Phys 4, 228}\ } (\bibinfo {year}
  {2021}),\ https://doi.org/10.1038/s42005-021-00730-0}\BibitemShut {NoStop}%
\bibitem [{\citenamefont {Wu}\ and\ \citenamefont
  {Lidar}(2002)}]{parafermions}%
  \BibitemOpen
  \bibfield  {author} {\bibinfo {author} {\bibfnamefont {L.-A.}\ \bibnamefont
  {Wu}}\ and\ \bibinfo {author} {\bibfnamefont {D.}~\bibnamefont {Lidar}},\
  }\href {\doibase https://doi.org/10.1063/1.1499208} {\bibfield  {journal}
  {\bibinfo  {journal} {Journal of Mathematical Physics}\ }\textbf {\bibinfo
  {volume} {43}},\ \bibinfo {pages} {4506} (\bibinfo {year}
  {2002})}\BibitemShut {NoStop}%
\bibitem [{\citenamefont {Yordanov}\ \emph {et~al.}(2020)\citenamefont
  {Yordanov}, \citenamefont {Arvidsson-Shukur},\ and\ \citenamefont
  {Barnes}}]{eff_circs}%
  \BibitemOpen
  \bibfield  {author} {\bibinfo {author} {\bibfnamefont {Y.~S.}\ \bibnamefont
  {Yordanov}}, \bibinfo {author} {\bibfnamefont {D.~R.~M.}\ \bibnamefont
  {Arvidsson-Shukur}}, \ and\ \bibinfo {author} {\bibfnamefont {C.~H.~W.}\
  \bibnamefont {Barnes}},\ }\href {\doibase 10.1103/PhysRevA.102.062612}
  {\bibfield  {journal} {\bibinfo  {journal} {Phys. Rev. A}\ }\textbf {\bibinfo
  {volume} {102}},\ \bibinfo {pages} {062612} (\bibinfo {year}
  {2020})}\BibitemShut {NoStop}%
\bibitem [{\citenamefont {Xia}\ and\ \citenamefont {Kais}(2020)}]{UCCSD_qubit}%
  \BibitemOpen
  \bibfield  {author} {\bibinfo {author} {\bibfnamefont {R.}~\bibnamefont
  {Xia}}\ and\ \bibinfo {author} {\bibfnamefont {S.}~\bibnamefont {Kais}},\
  }\href {\doibase https://doi.org/10.1088/2058-9565/abbc74} {\bibfield
  {journal} {\bibinfo  {journal} {Quantum Science and Technology}\ } (\bibinfo
  {year} {2020}),\ https://doi.org/10.1088/2058-9565/abbc74}\BibitemShut
  {NoStop}%
\bibitem [{\citenamefont {Wigner}\ and\ \citenamefont
  {Jordan}(1928)}]{JW_encode}%
  \BibitemOpen
  \bibfield  {author} {\bibinfo {author} {\bibfnamefont {E.}~\bibnamefont
  {Wigner}}\ and\ \bibinfo {author} {\bibfnamefont {P.}~\bibnamefont
  {Jordan}},\ }\href
  {https://link.springer.com/chapter/10.1007/978-3-662-02781-3_9} {\bibfield
  {journal} {\bibinfo  {journal} {Z. Phys}\ }\textbf {\bibinfo {volume} {47}},\
  \bibinfo {pages} {631} (\bibinfo {year} {1928})}\BibitemShut {NoStop}%
\bibitem [{\citenamefont {Bravyi}\ and\ \citenamefont
  {Kitaev}(2002)}]{BK_encode}%
  \BibitemOpen
  \bibfield  {author} {\bibinfo {author} {\bibfnamefont {S.~B.}\ \bibnamefont
  {Bravyi}}\ and\ \bibinfo {author} {\bibfnamefont {A.~Y.}\ \bibnamefont
  {Kitaev}},\ }\href {\doibase https://doi.org/10.1006/aphy.2002.6254}
  {\bibfield  {journal} {\bibinfo  {journal} {Annals of Physics}\ }\textbf
  {\bibinfo {volume} {298}},\ \bibinfo {pages} {210} (\bibinfo {year}
  {2002})}\BibitemShut {NoStop}%
\bibitem [{\citenamefont {McClean}\ \emph {et~al.}(2020)\citenamefont
  {McClean}, \citenamefont {Jiang}, \citenamefont {Rubin}, \citenamefont
  {Babbush},\ and\ \citenamefont {Neven}}]{subspace_expansion_2}%
  \BibitemOpen
  \bibfield  {author} {\bibinfo {author} {\bibfnamefont {J.~R.}\ \bibnamefont
  {McClean}}, \bibinfo {author} {\bibfnamefont {Z.}~\bibnamefont {Jiang}},
  \bibinfo {author} {\bibfnamefont {N.~C.}\ \bibnamefont {Rubin}}, \bibinfo
  {author} {\bibfnamefont {R.}~\bibnamefont {Babbush}}, \ and\ \bibinfo
  {author} {\bibfnamefont {H.}~\bibnamefont {Neven}},\ }\href {\doibase
  https://doi.org/10.1038/s41467-020-14341-w} {\bibfield  {journal} {\bibinfo
  {journal} {Nature Communications}\ }\textbf {\bibinfo {volume} {11}},\
  \bibinfo {pages} {1} (\bibinfo {year} {2020})}\BibitemShut {NoStop}%
\bibitem [{\citenamefont {McClean}\ \emph {et~al.}(2017)\citenamefont
  {McClean}, \citenamefont {Kimchi-Schwartz}, \citenamefont {Carter},\ and\
  \citenamefont {de~Jong}}]{subspace_expansion_3}%
  \BibitemOpen
  \bibfield  {author} {\bibinfo {author} {\bibfnamefont {J.~R.}\ \bibnamefont
  {McClean}}, \bibinfo {author} {\bibfnamefont {M.~E.}\ \bibnamefont
  {Kimchi-Schwartz}}, \bibinfo {author} {\bibfnamefont {J.}~\bibnamefont
  {Carter}}, \ and\ \bibinfo {author} {\bibfnamefont {W.~A.}\ \bibnamefont
  {de~Jong}},\ }\href {\doibase 10.1103/PhysRevA.95.042308} {\bibfield
  {journal} {\bibinfo  {journal} {Phys. Rev. A}\ }\textbf {\bibinfo {volume}
  {95}},\ \bibinfo {pages} {042308} (\bibinfo {year} {2017})}\BibitemShut
  {NoStop}%
\bibitem [{\citenamefont {Colless}\ \emph {et~al.}(2018)\citenamefont
  {Colless}, \citenamefont {Ramasesh}, \citenamefont {Dahlen}, \citenamefont
  {Blok}, \citenamefont {Kimchi-Schwartz}, \citenamefont {McClean},
  \citenamefont {Carter}, \citenamefont {de~Jong},\ and\ \citenamefont
  {Siddiqi}}]{subspace_expansion_1}%
  \BibitemOpen
  \bibfield  {author} {\bibinfo {author} {\bibfnamefont {J.~I.}\ \bibnamefont
  {Colless}}, \bibinfo {author} {\bibfnamefont {V.~V.}\ \bibnamefont
  {Ramasesh}}, \bibinfo {author} {\bibfnamefont {D.}~\bibnamefont {Dahlen}},
  \bibinfo {author} {\bibfnamefont {M.~S.}\ \bibnamefont {Blok}}, \bibinfo
  {author} {\bibfnamefont {M.~E.}\ \bibnamefont {Kimchi-Schwartz}}, \bibinfo
  {author} {\bibfnamefont {J.~R.}\ \bibnamefont {McClean}}, \bibinfo {author}
  {\bibfnamefont {J.}~\bibnamefont {Carter}}, \bibinfo {author} {\bibfnamefont
  {W.~A.}\ \bibnamefont {de~Jong}}, \ and\ \bibinfo {author} {\bibfnamefont
  {I.}~\bibnamefont {Siddiqi}},\ }\href {\doibase 10.1103/PhysRevX.8.011021}
  {\bibfield  {journal} {\bibinfo  {journal} {Phys. Rev. X}\ }\textbf {\bibinfo
  {volume} {8}},\ \bibinfo {pages} {011021} (\bibinfo {year}
  {2018})}\BibitemShut {NoStop}%
\bibitem [{\citenamefont {Higgott}\ \emph {et~al.}(2019)\citenamefont
  {Higgott}, \citenamefont {Wang},\ and\ \citenamefont
  {Brierley}}]{overlap_method_1}%
  \BibitemOpen
  \bibfield  {author} {\bibinfo {author} {\bibfnamefont {O.}~\bibnamefont
  {Higgott}}, \bibinfo {author} {\bibfnamefont {D.}~\bibnamefont {Wang}}, \
  and\ \bibinfo {author} {\bibfnamefont {S.}~\bibnamefont {Brierley}},\ }\href
  {\doibase https://doi.org/10.22331/q-2019-07-01-156} {\bibfield  {journal}
  {\bibinfo  {journal} {Quantum}\ }\textbf {\bibinfo {volume} {3}},\ \bibinfo
  {pages} {156} (\bibinfo {year} {2019})}\BibitemShut {NoStop}%
\bibitem [{\citenamefont {Jones}\ \emph {et~al.}(2019)\citenamefont {Jones},
  \citenamefont {Endo}, \citenamefont {McArdle}, \citenamefont {Yuan},\ and\
  \citenamefont {Benjamin}}]{overlap_method_2}%
  \BibitemOpen
  \bibfield  {author} {\bibinfo {author} {\bibfnamefont {T.}~\bibnamefont
  {Jones}}, \bibinfo {author} {\bibfnamefont {S.}~\bibnamefont {Endo}},
  \bibinfo {author} {\bibfnamefont {S.}~\bibnamefont {McArdle}}, \bibinfo
  {author} {\bibfnamefont {X.}~\bibnamefont {Yuan}}, \ and\ \bibinfo {author}
  {\bibfnamefont {S.~C.}\ \bibnamefont {Benjamin}},\ }\href {\doibase
  10.1103/PhysRevA.99.062304} {\bibfield  {journal} {\bibinfo  {journal} {Phys.
  Rev. A}\ }\textbf {\bibinfo {volume} {99}},\ \bibinfo {pages} {062304}
  (\bibinfo {year} {2019})}\BibitemShut {NoStop}%
\bibitem [{\citenamefont {Cincio}\ \emph {et~al.}(2018)\citenamefont {Cincio},
  \citenamefont {Suba{\c{s}}{\i}}, \citenamefont {Sornborger},\ and\
  \citenamefont {Coles}}]{swap_test}%
  \BibitemOpen
  \bibfield  {author} {\bibinfo {author} {\bibfnamefont {L.}~\bibnamefont
  {Cincio}}, \bibinfo {author} {\bibfnamefont {Y.}~\bibnamefont
  {Suba{\c{s}}{\i}}}, \bibinfo {author} {\bibfnamefont {A.~T.}\ \bibnamefont
  {Sornborger}}, \ and\ \bibinfo {author} {\bibfnamefont {P.~J.}\ \bibnamefont
  {Coles}},\ }\href {\doibase https://doi.org/10.1088/1367-2630/aae94a}
  {\bibfield  {journal} {\bibinfo  {journal} {New Journal of Physics}\ }\textbf
  {\bibinfo {volume} {20}},\ \bibinfo {pages} {113022} (\bibinfo {year}
  {2018})}\BibitemShut {NoStop}%
\bibitem [{\citenamefont {Kokail}\ \emph {et~al.}(2019)\citenamefont {Kokail},
  \citenamefont {Maier}, \citenamefont {van Bijnen}, \citenamefont {Brydges},
  \citenamefont {Joshi}, \citenamefont {Jurcevic}, \citenamefont {Muschik},
  \citenamefont {Silvi}, \citenamefont {Blatt}, \citenamefont {Roos} \emph
  {et~al.}}]{direct_search}%
  \BibitemOpen
  \bibfield  {author} {\bibinfo {author} {\bibfnamefont {C.}~\bibnamefont
  {Kokail}}, \bibinfo {author} {\bibfnamefont {C.}~\bibnamefont {Maier}},
  \bibinfo {author} {\bibfnamefont {R.}~\bibnamefont {van Bijnen}}, \bibinfo
  {author} {\bibfnamefont {T.}~\bibnamefont {Brydges}}, \bibinfo {author}
  {\bibfnamefont {M.~K.}\ \bibnamefont {Joshi}}, \bibinfo {author}
  {\bibfnamefont {P.}~\bibnamefont {Jurcevic}}, \bibinfo {author}
  {\bibfnamefont {C.~A.}\ \bibnamefont {Muschik}}, \bibinfo {author}
  {\bibfnamefont {P.}~\bibnamefont {Silvi}}, \bibinfo {author} {\bibfnamefont
  {R.}~\bibnamefont {Blatt}}, \bibinfo {author} {\bibfnamefont {C.~F.}\
  \bibnamefont {Roos}},  \emph {et~al.},\ }\href {\doibase
  doi.org/10.1038/s41586-019-1177-4} {\bibfield  {journal} {\bibinfo  {journal}
  {Nature}\ }\textbf {\bibinfo {volume} {569}},\ \bibinfo {pages} {355}
  (\bibinfo {year} {2019})}\BibitemShut {NoStop}%
\bibitem [{\citenamefont {Nelder}\ and\ \citenamefont
  {Mead}(1965)}]{nelder_mead}%
  \BibitemOpen
  \bibfield  {author} {\bibinfo {author} {\bibfnamefont {J.~A.}\ \bibnamefont
  {Nelder}}\ and\ \bibinfo {author} {\bibfnamefont {R.}~\bibnamefont {Mead}},\
  }\href {\doibase 10.1093/comjnl/7.4.308} {\bibfield  {journal} {\bibinfo
  {journal} {The Computer Journal}\ }\textbf {\bibinfo {volume} {7}},\ \bibinfo
  {pages} {308} (\bibinfo {year} {1965})},\ \Eprint
  {http://arxiv.org/abs/https://academic.oup.com/comjnl/article-pdf/7/4/308/1013182/7-4-308.pdf}
  {https://academic.oup.com/comjnl/article-pdf/7/4/308/1013182/7-4-308.pdf}
  \BibitemShut {NoStop}%
\bibitem [{\citenamefont {Ditchfield}\ \emph {et~al.}(1971)\citenamefont
  {Ditchfield}, \citenamefont {Hehre},\ and\ \citenamefont {Pople}}]{sto_3g}%
  \BibitemOpen
  \bibfield  {author} {\bibinfo {author} {\bibfnamefont {R.}~\bibnamefont
  {Ditchfield}}, \bibinfo {author} {\bibfnamefont {W.~J.}\ \bibnamefont
  {Hehre}}, \ and\ \bibinfo {author} {\bibfnamefont {J.~A.}\ \bibnamefont
  {Pople}},\ }\href {\doibase 10.1063/1.1674902} {\bibfield  {journal}
  {\bibinfo  {journal} {The Journal of Chemical Physics}\ }\textbf {\bibinfo
  {volume} {54}},\ \bibinfo {pages} {724} (\bibinfo {year} {1971})}\BibitemShut
  {NoStop}%
\bibitem [{\citenamefont {Fletcher}(2013)}]{bfgs}%
  \BibitemOpen
  \bibfield  {author} {\bibinfo {author} {\bibfnamefont {R.}~\bibnamefont
  {Fletcher}},\ }\href {https://archive.org/details/practicalmethods0000flet}
  {\emph {\bibinfo {title} {Practical methods of optimization}}}\ (\bibinfo
  {publisher} {John Wiley \& Sons},\ \bibinfo {year} {2013})\BibitemShut
  {NoStop}%
\bibitem [{\citenamefont {Helgaker}\ \emph {et~al.}(2012)\citenamefont
  {Helgaker}, \citenamefont {Coriani}, \citenamefont {Jorgensen}, \citenamefont
  {Kristensen}, \citenamefont {Olsen},\ and\ \citenamefont
  {Ruud}}]{classical_methods}%
  \BibitemOpen
  \bibfield  {author} {\bibinfo {author} {\bibfnamefont {T.}~\bibnamefont
  {Helgaker}}, \bibinfo {author} {\bibfnamefont {S.}~\bibnamefont {Coriani}},
  \bibinfo {author} {\bibfnamefont {P.}~\bibnamefont {Jorgensen}}, \bibinfo
  {author} {\bibfnamefont {K.}~\bibnamefont {Kristensen}}, \bibinfo {author}
  {\bibfnamefont {J.}~\bibnamefont {Olsen}}, \ and\ \bibinfo {author}
  {\bibfnamefont {K.}~\bibnamefont {Ruud}},\ }\href
  {https://pubs.acs.org/doi/pdf/10.1021/cr2002239} {\bibfield  {journal}
  {\bibinfo  {journal} {Chemical reviews}\ }\textbf {\bibinfo {volume} {112}},\
  \bibinfo {pages} {543} (\bibinfo {year} {2012})}\BibitemShut {NoStop}%
\end{thebibliography}%

\end{document}